\documentclass[aps, prb, superscriptaddress, reprint, twocolumn, a4, notitlepage]{revtex4-2}
\usepackage{graphicx}
\usepackage{amssymb}
\usepackage{amsmath,bm}
\usepackage{dcolumn}
\usepackage{mathrsfs}
\usepackage{physics}
\usepackage[utf8]{inputenc}
\usepackage{mathptmx}
\DeclareMathAlphabet{\mathcal}{OMS}{cmsy}{m}{n}
\newcommand{\widebar}[1]{\mkern 1.5mu\overline{\mkern-1.5mu#1\mkern-1.5mu}\mkern 1.5mu}

\usepackage{color}

\newcommand{\bk}{\mathbf{k}}

\newcommand{\bq}{\mathbf{q}}

\newcommand{\rev}[1]{#1}

\begin{document}
\title{Bichromatic four-wave mixing and quadrature-squeezing from biexcitons in atomically thin semiconductor microcavities}
\date{\today}

\author{Emil V. Denning}
\affiliation{Nichtlineare Optik und Quantenelektronik, Institut f\"ur Theoretische Physik, Technische Universit\"at Berlin, 10623 Berlin, Germany}

\author{Andreas Knorr}
\affiliation{Nichtlineare Optik und Quantenelektronik, Institut f\"ur Theoretische Physik, Technische Universit\"at Berlin, 10623 Berlin, Germany}

\author{Florian Katsch}
\affiliation{Nichtlineare Optik und Quantenelektronik, Institut f\"ur Theoretische Physik, Technische Universit\"at Berlin, 10623 Berlin, Germany}

\author{Marten Richter}
\affiliation{Nichtlineare Optik und Quantenelektronik, Institut f\"ur Theoretische Physik, Technische Universit\"at Berlin, 10623 Berlin, Germany}

\date{\today}

\begin{abstract}
Nonlinear optical effects such as four-wave mixing and generation of squeezed light are ubiquitous in optical devices and light sources. For new devices operating at low optical power, the resonant nonlinearity arising from the two-photon sensitive bound biexciton in a semiconductor microcavity is an interesting prospective platform. Due to the particularly strong Coulomb interaction in atomically thin semiconductors, these materials have strongly bound biexcitons and operate in the visible frequency range of the electromagnetic spectrum. To remove the strong pump laser from the generated light in optical devices or to simultaneously excite non-degenerate polaritons, a bichromatic-pump configuration with two spectrally separated pump lasers is desirable. In this paper, we theoretically investigate spontanous four-wave mixing and quadrature-squeezing in a bichromatically pumped atomically thin semiconductor microcavity. We explore two different configurations that support degenerate and non-degenerate scattering from polaritons into bound biexcitons, respectively. We find that these configurations lead to the generation strongly single- and two-mode quadrature-squeezed light.
\end{abstract}
\maketitle

\section{Introduction}

Strongly bound biexcitons in atomically thin semiconductor microcavities provide an avenue for low-power nonlinear optical devices, because the resonant scattering of unbound exciton-polaritons into a bound biexciton yields a powerful enhancement of the nonlinear optical response~\cite{wouters2007resonant,carusotto2010feshbach,camacho2022strong}. Previous analyses of degenerate four-wave mixing have shown that strong parametric gain from biexcitons~\cite{shimano2002efficient,savasta2003many} can provide quadrature-squeezing with significantly lower power than when using conventional third-order nonlinear materials~\cite{denning2022efficient}. Thereby, atomically-thin semiconductors have a significant potential as a platform for nonlinear optics. In the interest of spectrally separating the pump photons from the generated signal, it is highly convenient to use a bichromatic-pump scheme, where two non-degenerate pumps create a degenerate signal at the mean of the two pump photon energies~\cite{fang2013state,okawachi2015dual,he2015ultracompact,vernon2019scalable,zhang2021squeezed,seifoory2022degenerate}. In addition, such a bichromatic-pump setup allows to excite non-degenerate polaritons, which opens up a non-degenerate nonlinear scattering channel~\cite{navadeh2019polaritonic}.

\begin{figure}
  \centering
  \includegraphics[width=\columnwidth]{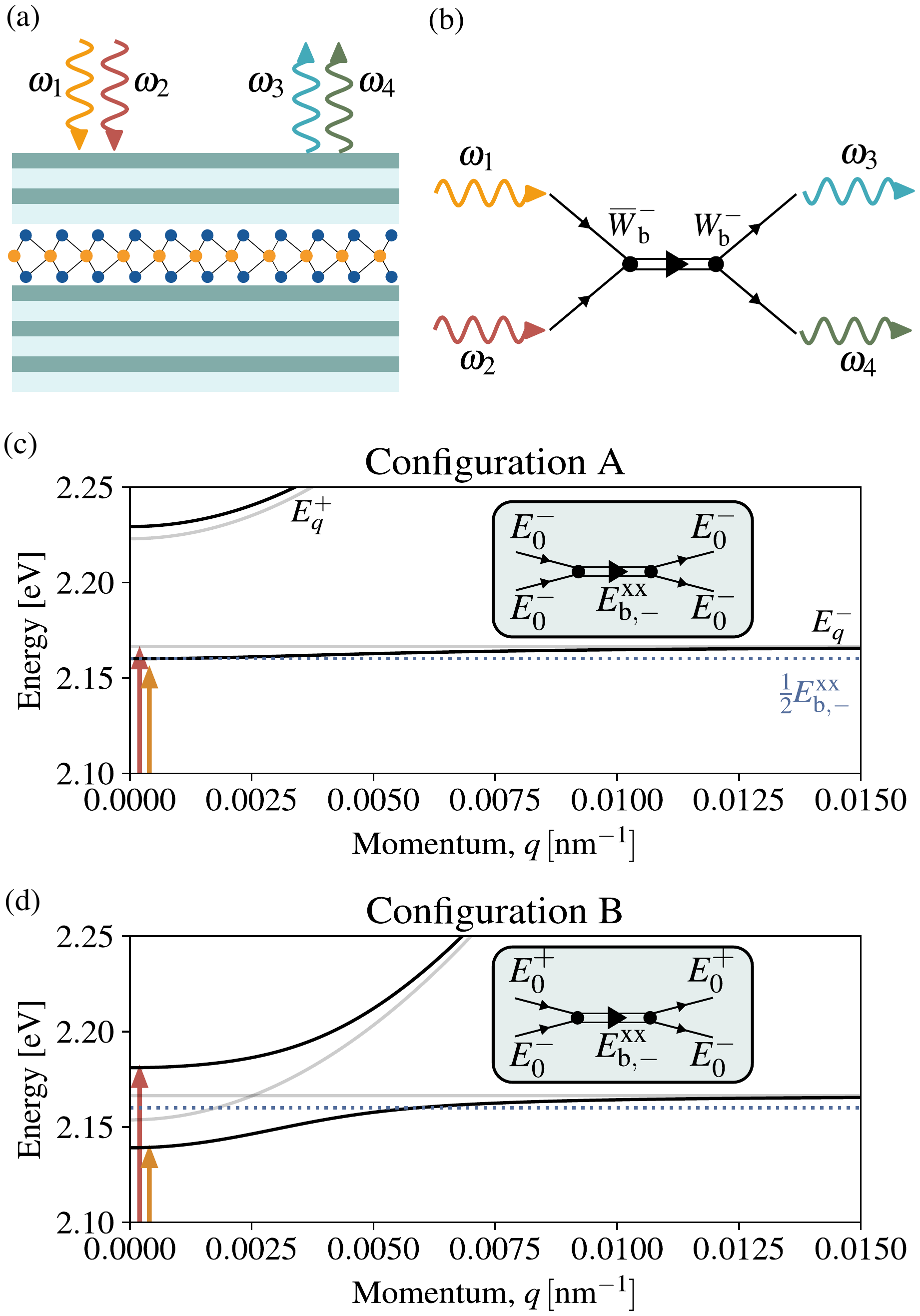}
  \caption{(a) Microcavity containing an atomically-thin semiconductor. The cavity is driven with bichromatic coherent laser light at frequencies $\omega_1$ and $\omega_2$. Four-wave mixing due to nonlinearities in the optical response of the semiconductor creates an output photon pair with frequencies $\omega_3$ and $\omega_4$. (b) Diagrammatic illustration of biexcitonic four-wave mixing. The input photons (red and yelow wiggly arrows) create a pair of uncorrelated exciton-polaritons (single black lines). The two polaritons can scatter via the Coulomb interaction $\overline{W}_{\rm b}^{-}$ and create a bound biexciton (double line). The biexciton spontaneously breaks into a pair of polaritons via the Coulomb interaction $W^{-}_{\rm b}$, which are outcoupled from the cavity as photons (blue and green wiggly arrows). (c) Polariton energy bands ($E^\pm_q$, black lines) in \emph{Configuration A}, where the cavity frequency is tuned in order to bring the lower polariton branch into Feshbach resonance with the bound biexciton (blue dotted line), $2E_0^-=E^{\rm xx}_{\rm b,-}$. The driving frequencies are symmetrically centered around the lower polariton (orange and red arrows). The uncoupled cavity and exciton energy bands are shown with grey lines. (d) Similar to (c) for \emph{Configuration B}, where the cavity frequency is tuned such that an upper-lower polariton pair matches the energy of a bound biexciton, $E^-_0 + E^+_0=E^{\rm xx}_{\rm b,-}$, forming a non-degenerate Feshbach resonance.
  The driving frequencies are symmetrically centered around $\frac{1}{2}E^{\rm xx}_{\rm b,-}$ with a resonance occurring when $\omega_1=E^+_0,\;\omega_2=E^-_0$.
  }
  \label{fig:1}
\end{figure}

Four-wave mixing in pump-probe experiments with semiconductors~\cite{abram1984nonlinear,mayer1994evidence,kuroda1995coherent,schafer1996femtosecond,akiyama1999biexcitonic,langbein2001spectral,katsch2019theory} and semiconductor microcavities~\cite{kuwata1997parametric,svirko2000signatures} is a heavily studied theoretical and experimental topic, and it was early recognized that the bound biexciton plays a paramount role in four-wave mixing. However, spontaneous four-wave mixing, also known as parametric flourescence or hyper-raman scattering, in semiconductors is far less studied~\cite{savasta1999hyper,kira1999quantumB}, despite being a central topic for quantum light sources with conventional nonlinear materials~\cite{harris1967observation,helt2010spontaneous,azzini2012classical,koefoed2019complete}. Four-wave mixing is a third-order nonlinear process that transforms two pump photons with frequencies $\omega_1,\;\omega_2$ to a pair of signal/idler photons with frequencies $\omega_3,\;\omega_4$. Stimulated four-wave mixing is facilitated by applying an idler laser field with frequency $\omega_3$, which will produce a signal at $\omega_4=\omega_1+\omega_2-\omega_3$ due to energy conservation. In contrast, in the absence of a stimulating idler field, spontaneous four-wave mixing produce a broad and continuous spectrum of photon pairs. This is the regime that is studied in this paper. Due to the pairwise creation of photon pairs, the field properties of the generated light can exhibit strong non-classical correlation signatures in the form of quadrature squeezing, two-mode squeezing and photon-number correlations. In semiconductors, particularly in semiconductor microcavities with a strong light-matter interaction, spontaneous four-wave mixing can be mediated by the biexciton, which can break into a correlated pair of polaritons [see Fig.~\ref{fig:1}(a)--(b)].

In this paper, we theoretically investigate spontaneous four-wave mixing in bichromatically pumped semiconductor microcavities with atomically-thin semiconductors. We investigate the spectrum and quadrature squeezing of the generated light and discuss how the cavity resonance and pump frequencies can be tuned in two configurations to facilitate efficient light generation and squeezing. Specifically, we investigate two configurations close to resonances, where four-wave mixing and squeezing is efficient. In the first configuration [\emph{Configuration A}, cf. Fig.~\ref{fig:1}(c)], the cavity frequency is tuned such that the lower polariton (LP) energy $E^-_0$ matches half the bound biexciton energy $\frac{1}{2}E^{\rm xx}_{\rm b,-}$. Here, the pump lasers should be nearly degenerate around the lower-polariton energy in order to efficiently excite LPs, which scatter via the Coulomb interaction into bound biexcitons through a degenerate polaritonic Feshbach resonance~\cite{wouters2007resonant,takemura2014polaritonic}. A Feshbach resonance occurs when the energy of an open scattering channel (in this case a LP pair) matches the energy of a bound multiparticle complex (here a bound biexciton)~\cite{chin2010feshbach}. The biexciton decays spontaneously into two degenerate LPs, which leads to single-mode squeezed light emission. In the second configuration [\emph{Configuration B}, cf. Fig.~\ref{fig:1}(d)], the cavity frequency is tuned such that the sum of the LP and upper polariton (UP) energies $E^-_0 + E^+_0$ matches the bound biexciton energy $E^{\rm xx}_{\rm b,-}$, thereby giving rise to a non-degenerate polaritonic Feshbach resonance, also known as a polaritonic cross Feshbach resonance~\cite{navadeh2019polaritonic}. The driving field is resonant, when the two pump laser energies match the two polariton energies. An uncorrelated UP-LP pair can then resonantly scatter into a bound biexciton, which decays into a non-degenerate UP-LP pair that leads to the emission of two-mode squeezed light. In this context, two-mode quadrature-squeezing refers to a reduction of the quadrature noise in the cross-correlation between spectrally distinct frequency bands~\cite{ekert1989correlations}.

For the investigation, we employ a rigorous perturbative expansion of the electronic and photonic correlations up to third order in the driving field through the dynamics-controlled truncation (DCT) scheme~\cite{axt1994dynamics,lindberg1994chi}. In order to calculate not only intracavity dynamics, but also the properties of the outcoupled and thus detectable field, we combine DCT with a Heisenberg-Langevin approach~\cite{portolan2008nonequilibrium,denning2022efficient}. Due to the presence of two non-degenerate pumps, the steady state of the system is not constant~\cite{de2007quantum}, but can be expressed as a discrete Fourier series. This series expansion complicates the solution of the Heisenberg-Langevin equations of motion in frequency space compared to the case of a constant steady-state, which we solve through a discrete Fourier-index formalism.

The paper is organized as follows: In Sec.~\ref{sec:model}, we describe the semiconductor model that is used for the electronic states in the semiconductor, the photonic states in the cavity and the external driving field. In Sec.~\ref{sec:methods}, the DCT and Heisenberg-Langevin methods are applied to calculate the dynamics, spontaneous four-wave mixing spectrum and squeezing. In Sec.~\ref{sec:results}, we present and discuss the results for Configurations A and B. Finally, we conclude in Sec.~\ref{sec:conclusion}.

\section{Model}
\label{sec:model}
Here, we describe the used two-band semiconductor model for electrons and holes, as well as the coupling to cavity photons and the external laser drive.
\subsection{Hamiltonian}
We consider an atomically thin semiconductor placed in a planar cavity, which is driven with two coherent pump beams with frequencies $\omega_1$ and $\omega_2$ at normal incidence. The electromagnetic field in the cavity is quantized through the bosonic annihilation and creation operators $a_{\sigma,\bk}$ and $a_{\sigma,\bk}^\dagger$, which describe cavity photons with polarizaton $\sigma$ and in-plane momentum $\bk$. We assume the cavity to be rotationally symmetric in the plane, such that the cavity mode is polarization-degenerate.

The semiconductor is described by electronic states in the conduction and valence bands with fermionic annihilation and creation operators $c_{\zeta\bk}, \; c_{\zeta\bk}^\dagger$ for the conduction band and $v_{\zeta\bk}, \; v_{\zeta\bk}^\dagger$ for the valence band. The index $\zeta=(\xi,s)$ labels valley ($\xi$) and spin ($s$). We shall restrict our analysis to the lowest-energy optical transitions in transition-metal dichalcogenide monolayers, which are located at the valleys $\xi=K$ and $\xi=K'$, respectively. Due to spin-orbit coupling, the lowest-energy transitions allow photons with right-hand ($\sigma=\mathrm{R}$) or left-hand ($\sigma=\mathrm{L}$) circular polarization to excite  electron-hole pairs with spin-valley combinations $\zeta=(K,\uparrow)$ and $\zeta=(K',\downarrow)$. Due to this unambiguous relation between photon polarization and electron spin and valley, we can absorb photon polarization into the index $\zeta$ and use the shorthand notation $\zeta\in\{K,K'\}$ to denote all three degrees of freedom.

The total Hamiltonian $H$ is divided into a noninteracting term $H_0$ and a Coulomb interaction term $H_{\rm C}$, such that $H=H_0+H_{\rm C}$. The noninteracting Hamiltonian describing electrons, photons and their coupling is given by
\begin{align}
\begin{split}
  H_0 &= \sum_{\zeta\bk} \qty[
  E^{\rm c}_\bk c_{\zeta,\bk}^\dagger c_{\zeta,\bk}
  +E^{\rm v}_\bk v_{\zeta,\bk}^\dagger v_{\zeta,\bk}
  + E^{\rm p}_\bk a_{\zeta,\bk}^\dagger a_{\zeta,\bk}
  ]
  \\
  & + \sum_{\zeta\bk\bq} \qty[A_\bq c_{\zeta,\bk+\bq}^\dagger v_{\zeta,\bk} a_{\zeta,\bq}
  + A_\bq^* a_{\zeta,\bq}^\dagger  v_{\zeta,\bk}^\dagger c_{\zeta,\bk+\bq}],
\end{split}
\end{align}
where the first two terms account for the energy of electrons in the conduction and valence band, $E^{\rm c}_\bk=E^{\rm c}_{0} + \hbar^2k^2/(2m_{\rm e})$ and $E^{\rm v}_\bk=E^{\rm v}_{0} - \hbar^2k^2/(2m_{\rm h})$, with $E^{\rm c}_0-E^{\rm v}_0$ the quasiparticle bandgap and $m_{\rm e}, \:m_{\rm h}$ the effective electron and hole masses. The third term accounts for the energy of cavity photons, $E^{\rm p}_\bk = \hbar[\omega_{\rm p,0}^2 + (ck/\bar{n})^2]^{1/2}$, where $\omega_{\rm p,0}$ is the resonance frequency at $\bk=0$, $c$ is the speed of light and $\bar{n}$ is the effective refractive index of the cavity mode~\cite{skolnick1998strong,osgood2021dielectric}. The last two terms describe electron-hole-photon coupling with strength $A_\bq=\sqrt{E^{\rm p}_0/E^{\rm p}_\bq}A_0$, which depends on the out-of-plane confinement of the cavity mode and the Bloch momentum matrix element of the semiconductor~\cite{denning2022quantum, denning2022cavity}.

The Coulomb interaction is given by~\cite{katsch2018theory}
\begin{align}
\label{eq:H-coulomb}
\begin{split}
  &H_{\rm C} = \frac{1}{2}\sum_{\bk_1\bk_2\bq}  \sum_{\zeta_1\zeta_2} V_\bq\Big(c^\dagger_{\zeta_1,\bk_1+\bq}c^\dagger_{\zeta_2,\bk_2-\bq}
  c_{\zeta_2,\bk_2}c_{\zeta_1,\bk_1}
  \\
  &+ v^\dagger_{\zeta_1,\bk_1+\bq}v^\dagger_{\zeta_2,\bk_2-\bq}
  v_{\zeta_2,\bk_2}v_{\zeta_1,\bk_1}
 + 2c^\dagger_{\zeta_1,\bk_1+\bq}v^\dagger_{\zeta_2,\bk_2-\bq}
  v_{\zeta_2,\bk_2}c_{\zeta_1,\bk_1}
  \Big).
\end{split}
\end{align}
Here, $V_\bq= e_0^2[2S\epsilon_0\epsilon_\bq q]^{-1}$ is the screened 2D Coulomb potential, where $e_0$ is the elementary charge, $S$ is the quantization surface area, $\epsilon_0$ is the vacuum permittivity, and $\epsilon_\bq$ is the dielectric function for 2D semiconductors which is described in Appendix~\ref{sec:appendix-numerical-calculations}. The inter- and intra-valley Coulomb exchange interaction has been neglected here, because in transition-metal dichalcogenides it is significantly weaker~\cite{katsch2018theory,katsch2019theory} than the direct interaction in Eq.~\eqref{eq:H-coulomb}. We note that the exchange interaction leads to effects such as biexciton fine structure~\cite{steinhoff2020dynamical} and a splitting of the exciton into branches with linear and quadratic dispersion~\cite{qiu2015nonanalyticity} that coincide at zero center-of-mass momentum. While these effects are rich and interesting, we consider here only the physics of the dominating nonlinear response from the direct Coulomb interaction and leave the additional inclusion of exchange effects to more detailed future analyses.

\subsection{Driving field}
\label{sec:driving-field}

The two pump laser drives are introduced via the input-output formalism~\cite{steck2007quantum,gardiner2004quantum}, which is based on the microscopic interaction between the internal cavity mode with the quantized continuum of external modes. By formally solving the equation of motion of the external field operators, the external input field is linked to the equation of motion for the cavity field operator as
\begin{align}
\label{eq:cavity-H-L-1}
  -i\hbar\partial_t a_{\zeta,0}^\dagger = [H,a_{\zeta,0}^\dagger] +i\hbar\gamma^{\rm p}a_{\zeta,0}^\dagger + i\hbar\sqrt{2\gamma^{\rm p}} a^{\rm in\dagger}_{\zeta},
\end{align}
where the last two terms describe outcoupling from the cavity mode and incoupling of the driving field, respectively. We take the expectation value of the driving field to be in a coherent state and to have the bichromatic form
\begin{align}
  \ev*{a^{\rm in}_\zeta(t)} = \ev*{a^{\rm in}_\zeta}_1 e^{i\omega_1 t} + \ev*{a^{\rm in}_\zeta}_2 e^{i\omega_2 t},
\end{align}
where $\omega_1$ and $\omega_2$ are the frequencies of the two driving lasers and $\ev*{a^{\rm in}_\zeta}_1$ and $\ev*{a^{\rm in}_\zeta}_2$ are the corresponding amplitudes. We shall take the total power in each of the driving lasers to be equal and denote by $\bm{\lambda}^{\rm in}_{i}, \; i=1,2$ the polarization vector of the two drives in the circular basis. We then express the input field components as polarization vectors as $\ev{\mathbf{a}^{\rm in}}_i = [\ev{a^{\rm in}_K}_i, \ev{a^{\rm in}_{K'}}_i]^{\rm T}$, where the superscript T denotes transposition. The input-field vectors are related to the total driving power $\mathcal{P}_{\rm in}$ as~\cite{steck2007quantum} $\ev{\mathbf{a}^{\rm in}}_i = [\mathcal{P}/(2E^{\rm p}_0)]^{1/2} \bm{\lambda}^{\rm in}_i$.

\section{Methods}
\label{sec:methods}
Here, we introduce the DCT scheme for time evolution of the expectation values and the Heisenberg-Langevin equations for the fluctuation operators. The model and techniques are a generalization of the work presented in Ref.~\onlinecite{denning2022efficient} to bichromatic driving fields.

\subsection{Time evolution}
\label{sec:time-evolution}
To calculate the time evolution of the system, we apply the DCT scheme~\cite{axt1994dynamics,lindberg1994chi} to perturbatively expand the equations of motion for the coherent expectation values to third order in the driving field $a^{\rm in}_\zeta$. The first step in this procedure is the Heisenberg equation of motion for a general operator $Q$, $-i\hbar\partial_t \ev*{Q} = \ev*{[H,Q]}$ (see Ref.~\onlinecite{denning2022efficient} for the explicit derivation).

The Coulomb interaction and the fermionic commutation relations generate expectation values with an unequal number of conduction and valence band operators $c_{\zeta,\bk}$ and $v_{\zeta,\bk}$, such as $\ev*{c^\dagger_{\zeta,\bk+\bq} c^\dagger_{\zeta',\bk'-\bq} c_{\zeta',\bk'} v_{\zeta,\bk}}$. Here, conduction-band electron densities $c_{\zeta',\bk'-\bq}^\dagger c_{\zeta',\bk'}$ are expressed perturbatively in terms of electron-hole pair operators using a unit-operator expansion~\cite{ivanov1993self,katsch2018theory} $c_{\zeta',\bk'-\bq}^\dagger c_{\zeta',\bk'} = \sum_{\zeta_1\bk_1} c_{\zeta',\bk'-\bq}^\dagger v_{\zeta_1,\bk_1} v_{\zeta_1,\bk_1}^\dagger c_{\zeta',\bk'} + \mathcal{O}[(a^{\rm in })^4]$, and similarly for valence-band hole densities. In this expansion, the Hilbert space has been restricted to pairs of conduction band electrons and valence band holes~\cite{katsch2018theory}, which means that the effects of unpaired electrons or holes have not been accounted for. In this paper, this assumption is valid, because the only source of carriers is optical excitation through the driving field, which creates electrons and holes in pairs. However, in the presence of unpaired electrons or holes through e.g. n- or p-doping, which we do not consider in this paper, the pair expansion is no longer valid.

Within the DCT scheme, the third-order perturbative expansion corresponds to only keeping terms with up to three normal-ordered electron-hole pair or cavity photon operators~\cite{savasta1996quantum}, since the number of pair or photon operators corresponds to the order in the driving field. Such a perturbative expansion is possible because all correlations have their origin in the action of the external driving field~\cite{axt1994dynamics,lindberg1994chi}. Furthermore, in the coherent-response limit, which is considered here, the third-order expectation values are systematically factorized as $\ev{c^\dagger v c^\dagger v v^\dagger c} = \ev{c^\dagger v c^\dagger v}\ev*{v^\dagger c}$ and $\ev{a^\dagger c^\dagger v v^\dagger c} = \ev{a^\dagger c^\dagger v}\ev*{v^\dagger c}$. This third-order factorization is exact in the coherent regime~\cite{savasta1996quantum}, i.e. when the only source of electrons and holes is excitation with coherent light near the resonances of the system, and when incoherent scattering processes e.g. via phonons can be neglected. Phonon scattering is later included phenomenologically through an exciton dephasing rate, which is obtained from a separate self-consistent microscopic calculation~\cite{selig2016excitonic}. This means that the validity of our approach is limited to the regime where cavity outcoupling ($\gamma^{\rm p}$) dominates over exciton dephasing ($\gamma^{\rm x}$), such that polaritons will be outcoupled before significant dephasing and scattering takes place. In practise, this limits the theory to the low-temperature limit, where phonon dephasing is slow. For the calculations presented in this paper, the temperature is 30 K, which yields a phonon-induced dephasing rate below 1 meV, while the cavity outcoupling is 9 meV, consistent with fabricated dielectric microcavities with transition-metal dichalcogenide monolayers~\cite{liu2015strong}. Thus, there is a separation between these two time scales by an order of magnitude.

The electron-hole are expanded on the complete set of excitonic eigenstates as $\ev*{c^\dagger_{\zeta\bk}v_{\zeta\bk'}} = \sum_{i} \phi^{i*}_{\beta\bk+\alpha\bk'}\ev*{P^{i\dagger}_{\zeta,\bk-\bk'}}$, where $\phi^{i}_\bk$ is the $i$th exciton wavefunction (solution to the Wannier equation) in momentum space with zero-momentum energy $E^{\mathrm{x}}_{i,0}$ and $P^{i\dagger}_{\zeta,\bq}$ is the creation operator of the corresponding exciton with center-of-mass momentum $\bq$, and $\alpha=m_{\rm e}/(m_{\rm e} + m_{\rm h})$ and $\beta=m_{\rm h}/(m_{\rm e} + m_{\rm h})$. We restrict the analysis to the lowest-energy ($i=\mathrm{1s}$) exciton, which is energetically separated from the next excitonic state by hundreds of meV for atomically thin transition-metal dichalcogenides~\cite{mak2010atomically,ramasubramaniam2012large} and we shall thus drop the index $i$ from here onwards, meaning that the electronic states are projected onto the lowest exciton wavefunction. This is valid, when the excitation energy is far away from any of the neighbouring exciton states. In practise, the projection of electronic states onto the lowest-energy exciton wavefunction ensures an efficient numerical evaluation of the dynamics and exciton-exciton scattering coefficients.

Two-photon expectation values are partitioned into factorized parts and correlations, defined as $\mathcal{D}^{\zeta\zeta'}_{\bq} := \ev*{a^\dagger_{\zeta,\bq}a^\dagger_{\zeta',-\bq}} - \ev*{a^\dagger_{\zeta,\bq}}\ev*{a^\dagger_{\zeta',-\bq}}$.
Three-particle photon-electron-hole correlations are defined as $\ev*{a_{\zeta,\bq}^\dagger c^\dagger_{\zeta',\bk-\alpha\bq}v_{\zeta',\bk+\beta\bq}}^{\rm c} = \ev*{a_{\zeta,\bq}^\dagger c^\dagger_{\zeta',\bk-\alpha\bq}v_{\zeta',\bk+\beta\bq}} - \ev*{a_{\zeta,\bq}^\dagger}\!\!\ev*{ c^\dagger_{\zeta',\bk-\alpha\bq}v_{\zeta',\bk+\beta\bq}}$, i.e. with the contribution that can be factorized in electron-hole pairs and photons subtracted. These correlations are projected onto the exciton wavefunction as $\mathcal{C}^{\zeta\zeta'}_\bq := \sum_\bk \phi_{\bk}\ev*{a_{\zeta,\bq}^\dagger c^\dagger_{\zeta',\bk-\alpha\bq}v_{\zeta',\bk+\beta\bq}}^{\rm c}$. The four-particle correlations of two electron-hole pairs have a more complicated structure due to the two possible electron-hole pairings: $\ev*{c^\dagger_{\zeta\bk+\bq}v_{\zeta\bk}c^\dagger_{\zeta'\bk'-\bq}v_{\zeta'\bk'}}^{\rm c} := \ev*{c^\dagger_{\zeta\bk+\bq}v_{\zeta\bk}c^\dagger_{\zeta'\bk'-\bq}v_{\zeta'\bk'}} - \ev*{c^\dagger_{\zeta\bk+\bq}v_{\zeta\bk}}\!\!\ev*{c^\dagger_{\zeta'\bk'-\bq}v_{\zeta'\bk'}} + \ev*{c^\dagger_{\zeta'\bk'-\bq}v_{\zeta\bk}}\!\!\ev*{c^\dagger_{\zeta\bk+\bq}v_{\zeta'\bk'}}$. These correlations are projected on the 1s exciton wavefunction in terms of the biexcitonic  correlations $\tilde{\mathcal{B}}^{\zeta\zeta'}_{\bq,\pm}$ in the triplet ($+$) and singlet ($-$) linear combinations through the relation~\cite{schafer2013semiconductor}
\begin{align*}
&\frac{1}{2}\qty(\ev*{c_{\zeta\bk+\bq}^\dagger v_{\zeta\bk}
     c_{\zeta\bk'-\bq}^\dagger v_{\zeta'\bk'}}^{\rm c}
\pm
\ev*{c_{\zeta'\bk+\bq}^\dagger v_{\zeta\bk}
     c_{\zeta\bk'-\bq}^\dagger v_{\zeta'\bk'}}^{\rm c})
\\     &=:
\phi_{\bk + \beta\bq}^{*}
\phi_{\bk' - \beta\bq}^{*}
\tilde{\mathcal{B}}_{\bq,\pm}^{\zeta\zeta'}
\mp
\phi_{\alpha\bk + \beta(\bk'-\bq)}^{*}
\phi_{\beta(\bk+\bq) + \alpha\bk'}^{*}
\tilde{\mathcal{B}}_{\bk'-\bk-\bq,\pm}^{\zeta\zeta'}.
\end{align*}
Due to the Coulomb interaction, the equation of motion for the biexcitonic correlation $\tilde{\mathcal{B}}^{\zeta\zeta'}_{\bq,\pm}$ is coupled to correlations with different momenta, $\tilde{\mathcal{B}}^{\zeta\zeta'}_{\bq',\pm}$. To alleviate this momentum off-diagonal coupling and thereby simplify the structure of the equations of motion, the biexcitonic correlations are expanded on the biexcitonic wavefunctions $\Phi^{\pm}_{\mu,\bq}$, which are the solutions to an effective two-exciton Schr\"odinger equation~\cite{katsch2019theory} with corresponding energies $E^{\rm xx}_{\mu,\pm}$ (see Appendix~\ref{sec:matrix-elements} for details). Bound ($\mu=\mathrm{b},\:E_{\rm b,-}^{\mathrm{xx}}<2E_{0}^{\rm x}$) and unbound ($E_{\mu,-}^{{\rm xx}}>2E_{0}^{\rm x}$) solutions exist in the singlet channel, where the effective exciton-exciton Coulomb interaction is attractive. The triplet channel supports only unbound solutions~\cite{takayama2002T}, because the effective interaction is repulsive. The unbound solutions constitute a two-exciton scattering continuum. In this biexcitonic eigenbasis, the correlations are expressed as $\tilde{\mathcal{B}}^{\zeta\zeta'}_{\bq,\pm}=\sum_\mu \Phi^{\pm}_{\mu\bq} \mathcal{B}^{\zeta\zeta'}_{\mu,\pm}$.

At the level of the equations of motion, we phenomenologically include phonon-induced broadening of the exciton and biexciton by introducing complex-valued energies in the equations of motion: $\tilde{E}^{\rm x}_\bq=E^{\rm x}_\bq + i\hbar\gamma^{\rm x}, \: \tilde{E}^{\rm xx}_{\mu,\pm}=E^{\rm xx}_{\mu,\pm} + 2i\hbar\gamma^{\rm x}$. The broadening $\gamma^{\rm x}$ is calculated through a microscopic, self-consistent approach for the phonon interaction as in Refs.~\onlinecite{selig2016excitonic,christiansen2017phonon,khatibi2018impact,brem2019intrinsic,lengers2020theory}. We approximate the biexcitonic broadening as twice the exciton broadening~\cite{sieh1999coulomb,schumacher2005coherent,schumacher2006coherent}.

Similarly, photon outcoupling from the cavity is included in the complex cavity frequency $\tilde{E}^{\rm p}_\bq = E^{\rm p}_\bq + i\hbar\gamma^{\rm p}$.

The equations of motion of the photon and exciton amplitudes and the three types of correlations as described above form a closed set within the third-order DCT scheme~\cite{savasta1996quantum}. In a rotating reference frame with respect to the mean drive frequency $\omega_{\rm r}=(\omega_1+\omega_2)/2$ the equations of motion are
\begin{align}
\label{eq:eom}
\begin{split}
  -i\hbar\partial_t \ev*{a_{\zeta,0}^\dagger} &= (\tilde{E}^{\rm p}_0-\hbar\omega_{\rm r})\ev*{a_{\zeta,0}^\dagger} + \Omega_0 \ev*{P^\dagger_{\zeta,0}}
  \\
  &+i\hbar\sqrt{2\gamma^{\rm p}}[\ev*{a^{\rm in\dagger}_{\zeta}}_1e^{-i\omega_{12} t} + \ev*{a^{\rm in\dagger}_{\zeta}}_{2}e^{+i\omega_{12} t}]
  \\
  -i\hbar\partial_t\ev*{P_{\zeta,0}^\dagger} &= (\tilde{E}^{\rm x}_0-\hbar\omega_{\rm r})\ev*{P^{\dagger}_{\zeta,0}} + \Omega_0\ev*{a_{\zeta,0}^\dagger}
  \\ &-
  \sum_\bq \tilde{\Omega}_\bq\qty(\mathcal{C}^{\zeta\zeta'}_\bq + \delta_{\bq,0}\ev*{a_{\zeta,0}^\dagger}\ev*{P_{\zeta,0}^\dagger})\ev*{P_{\zeta,0}}
  \\ &+ W^0 \abs*{\ev*{P_{\zeta,0}^\dagger}}^2\ev*{P_{\zeta,0}^\dagger}
  +\sum_{\mu\zeta'\pm} W_{\mu}^{\pm}
  \mathcal{B}_{\mu,\pm}^{\zeta\zeta'} \ev*{P_{\zeta',0}}.
\\
-i\hbar\partial_t \mathcal{B}_{\mu,\pm}^{\zeta\zeta'} &= (\tilde{E}^{\rm xx}_{\mu,\pm}-2\hbar\omega_{\rm r})\mathcal{B}_{\mu,\pm}^{\zeta\zeta'}
+%\\&+
\frac{1}{2}(1\pm\delta_{\zeta\zeta'})
\\ &\hspace{-0.5cm}\times
\{\widebar{W}^\pm_{\mu}
\ev*{P^\dagger_{\zeta,0}}\ev*{P^\dagger_{\zeta',0}}
+
\sum_{\bq}[
\widebar{\Omega}_{\mu,-\bq}^\pm
\mathcal{C}^{\zeta'\zeta}_{-\bq}
+
\widebar{\Omega}_{\mu,\bq}^\pm
\mathcal{C}^{\zeta\zeta'}_\bq
]
\}
\\
-i\hbar\partial_t \mathcal{C}^{\zeta\zeta'}_\bq &=
(\tilde{E}^{\rm p}_{\bq}+\tilde{E}^{\rm x}_{\bq}
 - 2\hbar\omega_{\rm r})\mathcal{C}^{\zeta\zeta'}_\bq
 \\
 &+ \Omega_{\bq} \mathcal{D}^{\zeta\zeta'}_{\bq}
 - \frac{1}{2}\delta_{\zeta\zeta'}\tilde{\Omega}_\bq \ev*{P^\dagger_{\zeta,0}}^2
 + \sum_{\mu\pm} \Omega_{\mu,\bq}^{\pm}\mathcal{B}_{\mu,\pm}^{\zeta\zeta'}
\\
-i\hbar\partial_t \mathcal{D}^{\zeta\zeta'}_\bq &=
2(\tilde{E}^{\rm p}_\bq - \hbar\omega_{\rm r})
\mathcal{D}^{\zeta\zeta'}_\bq
+ \Omega_\bq \mathcal{C}^{\zeta'\zeta}_{-\bq}
+ \Omega_{-\bq}\mathcal{C}^{\zeta\zeta'}_\bq
.
\end{split}
\end{align}
The full details of the derivation can be found in Ref.~\onlinecite{denning2022efficient}, but for completeness, all quantities are defined in Appendix~\ref{sec:matrix-elements}.
In each equation, the first term on the right-hand side describes free evolution, and the remaining terms describe couplings or driving. For the photon amplitude $\ev*{a^\dagger_{\zeta,0}}$, the second term is the linear vacuum-Rabi coupling to the exciton with coupling strength $\Omega_0$ (where $2\Omega_0$ is the vacuum Rabi splitting). The last term is input-field driving, where in the rotating frame, the two pumps rotate with $\pm\omega_{12}$, with the difference frequency $\omega_{12}:=(\omega_2-\omega_1)/2$.

For $\ev*{P^\dagger_{\zeta,0}}$, the second term describes vacuum Rabi coupling. The third term arises from the fermionic substructure of excitons and generates nonlinear saturation of the light-matter interaction due to Pauli blocking $\tilde{\Omega}_\bq$. The last two terms describe Coulomb exciton-exciton interactions at the mean-field level ($W^0$) and Coulomb-induced interactions with the biexcitonic correlations ($W^\pm_\mu$) beyond mean-field.

For the biexcitonic correlations $\mathcal{B}^{\zeta\zeta'}_{\mu,\pm}$, the second term contains Coulomb-scattering of uncorrelated excitons ($\widebar{W}^\pm_\mu$) and coupling to exciton-photon correlations through the light-matter interaction ($\widebar{\Omega}_{\mu,\bq}^\pm$).
For the exciton-photon correlations $\mathcal{C}^{\zeta\zeta'}_\bq$, the second term describes linear coupling to two-photon correlations by exchanging an exciton with a photon ($\Omega_0$). The third term describes a nonlinear scattering of two uncorrelated excitons ($\tilde{\Omega}_\bq$), and the last term describes coupling to biexcitonic correlations via optical fields ($\Omega^\pm_{\mu,\bq}$). The second and third terms in the equation of motion for the two-photon correlations $\mathcal{D}^{\zeta\zeta'}_\bq$ describe coupling to exciton-photon correlations by exchanging a photon with an exciton through the light-matter coupling $\Omega_0$.

\subsection{Steady-state discrete Fourier decomposition}

As noted, the driving terms oscillate with the frequency $\pm\omega_{12}$ in the rotating frame. In contrast to the single-pump case~\cite{denning2022efficient} where the driving term is constant in the rotating frame, this means that the steady state evolution of the expectation values is not constant, but contains terms oscillating with integer multiples of $\omega_{12}$. This means that the steady-state dynamical variables $Y(t)$, where $Y$ is any of the expectation values in Eq.~\eqref{eq:eom}, are periodic with period $T=2\pi/\abs*{\omega_{12}}$ and can thus be represented as a discrete Fourier series
\begin{align}
\label{eq:discrete-fourier-definition}
\begin{split}
  Y(t)\big\vert_{t\rightarrow\infty} &= \sum_{m=-\infty}^\infty Y_m e^{-in\omega_{12}t}  \\
  Y_m &= \lim_{t\rightarrow\infty}\frac{1}{T}\int_{t}^{t+T}\dd{t'} Y(t')e^{im\omega_{12}t'}.
\end{split}
\end{align}
In order to find the steady-state discrete Fourier series, Eq.~\eqref{eq:eom} is propagated numerically in time until the coefficients $Y_m$ have converged sufficiently.

\subsection{Polaritons}
\label{sec:polaritons}
When photons and excitons are strongly coupled, as considered in this paper, the relevant linear-response eigenstates are polaritonic states. These states emerge naturally out of the equation of motion by diagonalization in the linear limit. Taking the first two lines of Eq.~\eqref{eq:eom}, removing the nonlinear terms and the driving, and casting them in the general nonzero momentum form in the laboratory (i.e. non-rotating) reference frame, we have
\begin{align}
\label{eq:linear-eom}
  \begin{split}
  -i\hbar\partial_t \ev*{a_{\zeta,\bq}^\dagger} &= \tilde{E}^{\rm p}_\bq\ev*{a_{\zeta,\bq}^\dagger} + \Omega_\bq \ev*{P^\dagger_{\zeta,\bq}}
  \\
  -i\hbar\partial_t\ev*{P_{\zeta,\bq}^\dagger} &= \tilde{E}^{\rm x}_\bq \ev*{P^{\dagger}_{\zeta,\bq}} + \Omega_\bq\ev*{a_{\zeta,\bq}^\dagger}.
  \end{split}
\end{align}
In the limit where photon outcoupling and exciton dephasing are neglected, these equations of motion can be diagonalized through a unitary Hopfield transformation~\cite{hopfield1958theory} by introducing the polariton operators $\Xi^{\pm\dagger}_{\zeta,\bq} = u^{\rm p\pm}_\bq a^\dagger_{\zeta,\bq} + u^{\rm x\pm}_\bq P^\dagger_{\zeta,\bq}$ where the expansion coefficients are given by
\begin{align}
\begin{split}
u^{\rm p\pm}_\bq &= \frac{E^{\rm p}_\bq - E^{\rm x}_\bq \pm \eta_\bq}{\sqrt{(E^{\rm p}_\bq - E^{\rm x}_\bq \pm \eta_\bq)^2 + 4\Omega_\bq^2}}
\\
u^{\rm x\pm}_\bq &= \frac{2\Omega_\bq}{\sqrt{(E^{\rm p}_\bq - E^{\rm x}_\bq \pm \eta_\bq)^2 + 4\Omega_\bq^2}},
\end{split}
\end{align}
with $\eta_\bq = \sqrt{(E^{\rm p}_\bq - E^{\rm x}_\bq)^2 + 4\Omega_\bq^2}$. In this basis, the linear evolution of Eq.~\eqref{eq:linear-eom} becomes
\begin{align}
  -i\hbar\partial_t \ev*{\Xi^{\pm\dagger}_{\zeta,\bq}} &= E^\pm_\bq\ev*{\Xi^{\pm\dagger}_{\zeta,\bq}},
\end{align}
with the polariton energies $E^\pm_\bq = \frac{1}{2}[E^{\rm x}_\bq + E^{\rm p}_\bq \pm \eta_\bq]$. The commutation relations of the polariton operators are not bosonic, but have correction terms due to the fermionic substructure of the exciton operator $P^\dagger_{\zeta,\bq}$~\cite{katsch2018theory}.
In principle, the full equations of motion Eq.~\eqref{eq:eom} can be expressed in the polaritonic basis. However, this is not necessary and will not change the dynamics itself, only the basis that it is expressed in. In this work, we only use the polaritonic energies to identify the linear resonances of the system and to understand the nonlinear scattering between zero-momentum polaritons and the bound biexciton as depicted in Fig.~\ref{fig:1}.

\subsection{Fluctuations}
\label{sec:fluctuations}

With the equations of motion of the single-time expectation values as in Eq.~\eqref{eq:eom}, one can access the instantaneous quantum statistics in the steady state or in the transient evolution. This can in principle be used to calculate the intracavity squeezing or photon number. However, the relevant detectable quantity is not the intracavity field, but the outcoupled field. Since the cavity field at zero-momentum couples out of the cavity into a continuum of external radiation modes with different frequencies, outcoupled quantities are always described and often measured by a spectrum rather than a single number, in contrast to intracavity quantities, which can be expressed in terms of a single mode~\cite{gardiner2004quantum,mork2020squeezing}.

To calculate the spontaneous four-wave mixing spectrum and quadrature squeezing of the generated light, we need not only the single-time expectation values as in Sec.~\ref{sec:time-evolution}, but also multitime averages. To calculate these, we employ a Heisenberg-Langevin approach for the time evolution of the fluctuation operator of the cavity field, $\delta a^{\dagger}_{\zeta,0}(t) = \lim_{t\rightarrow\infty}[a^{\dagger}_{\zeta,0}(t) - \ev*{a^{\dagger}_{\zeta,0}(t)}]$ and similarly for the exciton fluctuation operator $\delta P^\dagger_{\zeta,0}$. Importantly, the cavity in/out-coupling and the phonon-induced dephasing give rise to Langevin noise sources~\cite{lax1966quantum}. For the cavity field, the Langevin noise enters directly from microscopic theory through the input-output formalism, as seen in Eq.~\eqref{eq:cavity-H-L-1}, where $\delta a^{\rm in\dagger}_{\zeta}(t):=\lim_{t\rightarrow\infty}[a^{\rm in\dagger}_{\zeta}(t) - \ev*{a^{\rm in\dagger}_{\zeta}(t)}]$ is a Langevin noise term with the properties~\cite{gardiner2004quantum,steck2007quantum}
$
  \ev*{\delta a^{\rm in\dagger}_\zeta} = \ev*{\delta a^{\rm in\dagger}_\zeta(t)\delta a^{\rm in\dagger}_{\zeta'}(t')}=\ev*{\delta a^{\rm in\dagger}_\zeta(t)\delta a^{\rm in}_{\zeta'}(t')}=0,\;\;
  \ev*{\delta a^{\rm in}_\zeta(t)\delta a^{\rm in\dagger}_{\zeta'}(t')} = \delta_{\zeta\zeta'}\delta(t-t').
$
For the phenomenologically introduced phonon-broadening, we introduce a similar Langevin noise term to accompany the decay process,
\begin{align}
\label{eq:exciton-H-L-1}
  -i\hbar\partial_t P^\dagger_{\zeta,0} = [H,P^\dagger_{\zeta,0}] + i\hbar\gamma^{\rm x}P^\dagger_{\zeta,0}
  + i\hbar\sqrt{2\gamma^{\rm x}}\delta P^{\rm in\dagger}_{\zeta},
\end{align}
where the noise correlation properties are derived from the rules in Ref.~\onlinecite{lax1966quantum}
$
  \ev*{\delta P^{\rm in\dagger}_\zeta} = \ev*{\delta P^{\rm in\dagger}_\zeta(t)\delta P^{\rm in\dagger}_{\zeta'}(t')}=\ev*{\delta P^{\rm in\dagger}_\zeta(t)\delta P^{\rm in}_{\zeta'}(t')}=0,
  \ev*{\delta P^{\rm in}_\zeta(t)\delta P^{\rm in\dagger}_{\zeta'}(t')} = \delta_{\zeta\zeta'}\delta(t-t').
$

Calculating the commutator in Eq.~\eqref{eq:exciton-H-L-1} in a third-order expansion in the input field, it is found that the nonlinear terms couple $\delta P^\dagger_{\zeta,0}$ to exciton-photon pair fluctuation operators $\delta \mathcal{C}^{\zeta\zeta'}_{\bq}$ and biexcitonic fluctuations $\delta\mathcal{B}^{\zeta\zeta'}_{\mu,\pm}$ (see Ref.~\onlinecite{denning2022efficient} for details). For sufficiently small fluctuations compared to the mean values, the nonlinear terms involving products of fluctuations of the forms $\delta P\delta \mathcal{C}$,\; $\delta P\delta \mathcal{B}, \;\delta P^{\rm in\dagger}\delta P^\dagger, \; \delta a^{\rm in\dagger}\delta a^\dagger, \; \delta P^{\rm in\dagger}\delta a^\dagger$ and $\delta a^{\rm in\dagger}\delta P^\dagger$ can be neglected and a set of equations that are linear in the fluctuation operators is obtained,
\begin{align}
\begin{split}
\label{eq:fluctuation-eom-1}
  -i\hbar\partial_t\delta a^\dagger_{\zeta,0} &=
  (\tilde{E}^{\rm p}_0 - \hbar\omega_{\rm r}) \delta a^\dagger_{\zeta,0}
  + \Omega_0\delta P^\dagger_{\zeta,0} + i\hbar\sqrt{2\gamma^{\rm p}}\delta a^{\rm in\dagger}_{\zeta},
\\
  -i\hbar\partial_t \delta P_{\zeta,0}^\dagger
  &= (\tilde{E}_0^{\rm x}-\hbar\omega_{\rm r}) \delta P_{\zeta,0}^\dagger
  + \Omega_0\delta a_{\zeta,0}^\dagger
  + i\hbar\sqrt{2\gamma^{\rm x}} \delta P_{\zeta,0}^{\rm in\dagger}
  \\
  &\hspace{-1cm}-\sum_\bq \tilde{\Omega}_\bq
  \qty[
  \delta\mathcal{C}^{\zeta\zeta}_\bq
  \ev*{P^\dagger_{\zeta, 0}}
  +
  \qty(\delta_{\bq,0}\ev*{a_{\zeta,0}^\dagger}\ev*{P_{\zeta,0}^\dagger}
  + \mathcal{C}^{\zeta\zeta}_\bq)
  \delta P_{\zeta, 0}]
  \\
  &\hspace{-1cm}+W^0
  \ev*{P^\dagger_{\zeta,0}}^2
  \delta P_{\zeta,0}
  +
  \sum_{\zeta'\mu\pm}
  W^\pm_\mu
  \qty[\delta\mathcal{B}_{\mu,\pm}^{\zeta\zeta'}\ev*{P_{\zeta',0}}
  +\mathcal{B}_{\mu,\pm}^{\zeta\zeta'} \delta P_{\zeta',0}]
  \\
-i\hbar\partial_t \delta\mathcal{B}_{\mu,\pm}^{\zeta\zeta'} &= (\tilde{E}^{\rm xx}_{\mu,\pm}-2\hbar\omega_{\rm r})\delta\mathcal{B}_{\mu,\pm}^{\zeta\zeta'}
\\
&+
\frac{1}{2}(1\pm\delta_{\zeta\zeta'})
\qty[
\widebar{\Omega}_{\mu,-\bq}^\pm
\delta\mathcal{C}^{\zeta'\zeta}_{-\bq}
+
\widebar{\Omega}_{\mu,\bq}^\pm
\delta\mathcal{C}^{\zeta\zeta'}_\bq]
\\
&\hspace{-1cm}+
\frac{i\hbar}{4}\sqrt{2\gamma^{\rm x}}(1\pm\delta_{\zeta\zeta'})
\widebar{\Phi}^\pm_{\mu,0}
\qty[
\ev*{P^\dagger_{\zeta,0}}\delta P^{\rm in\dagger}_{\zeta'}
+ \ev*{P^\dagger_{\zeta',0}}\delta P^{\rm in\dagger}_{\zeta}]
\\
-i\hbar\partial_t\delta{\mathcal{C}}^{\zeta\zeta'}_{\bq} &=
(\tilde{E}^{\rm x}_\bq + \tilde{E}^{\rm p}_\bq-2\hbar\omega_{\rm r})\delta{\mathcal{C}}^{\zeta\zeta'}_{\bq}
+ \Omega_\bq\delta\mathcal{D}^{\zeta\zeta'}_\bq
\\
&+ \sum_{\mu\pm} \Omega^\pm_{\mu,\bq} \delta\mathcal{B}^{\zeta\zeta'}_{\mu,\pm}
+i\hbar\delta_{\bq,0}\sqrt{2\gamma^{\rm x}}\ev*{a^\dagger_{\zeta,0}}\delta P^{\rm in\dagger}_{\zeta'}
\\
&+i\hbar\delta_{\bq,0}\sqrt{2\gamma^{\rm p}}\qty[
\ev*{a^{\rm in\dagger}_{\zeta}}\delta P^\dagger_{\zeta',0}
+\ev*{P^\dagger_{\zeta',0}}\delta a^{\rm in\dagger}_{\zeta}]
\\
-i\hbar\partial_t\delta\mathcal{D}^{\zeta\zeta'}_\bq &=
2(\tilde{E}^{\rm p}-\hbar\omega_{\rm r})\delta\mathcal{D}^{\zeta\zeta'}_\bq
+ \Omega_\bq\delta\mathcal{C}^{\zeta'\zeta}_{-\bq}
+ \Omega_{-\bq}\delta\mathcal{C}^{\zeta\zeta'}_{\bq}
\\ &
+ i\hbar\sqrt{2\gamma^{\rm p}}\delta_{\bq,0}
\big[
\ev*{a^{\rm in\dagger}_{\zeta}}\delta a^\dagger_{\zeta',0}
+ \ev*{a^{\rm in\dagger}_{\zeta'}}\delta a^\dagger_{\zeta,0}
\\
&\hspace{2cm}+ \ev*{a^\dagger_{\zeta,0}}\delta a^{\rm in\dagger}_{\zeta'}
+ \ev*{a^\dagger_{\zeta',0}}\delta a^{\rm in\dagger}_{\zeta}
\big].
\end{split}
\end{align}

Being interested in the steady-state emission properties, we consider the limit of $t\rightarrow\infty$ and write all expectation values using their discrete Fourier series (see Eq.~\eqref{eq:discrete-fourier-definition}). We then transform the fluctuation equations to Fourier space as $\delta Q(\omega) = \int_{-\infty}^{\infty}\dd{t}e^{i\omega t}\delta Q(t)$, where $\delta Q$ is any of the fluctuation operators. Terms on the right-hand side of Eq.~\eqref{eq:fluctuation-eom-1} of the form $\delta Q(t)Y(t)$ transform as $\sum_m\delta Q(\omega-n\omega_{12})Y_m$, where $Y$ is a steady-state expectation value and  $Y_m$ is its discrete Fourier decomposition from Eq.~\eqref{eq:discrete-fourier-definition}. Thus, the fluctuation equations are not diagonal in frequency space, since $\delta Q(\omega)$ is coupled to $\delta Q(\omega+n\omega_{12})$ for $n\in\mathbb{Z}$.

\begin{figure}
\includegraphics[width=\columnwidth]{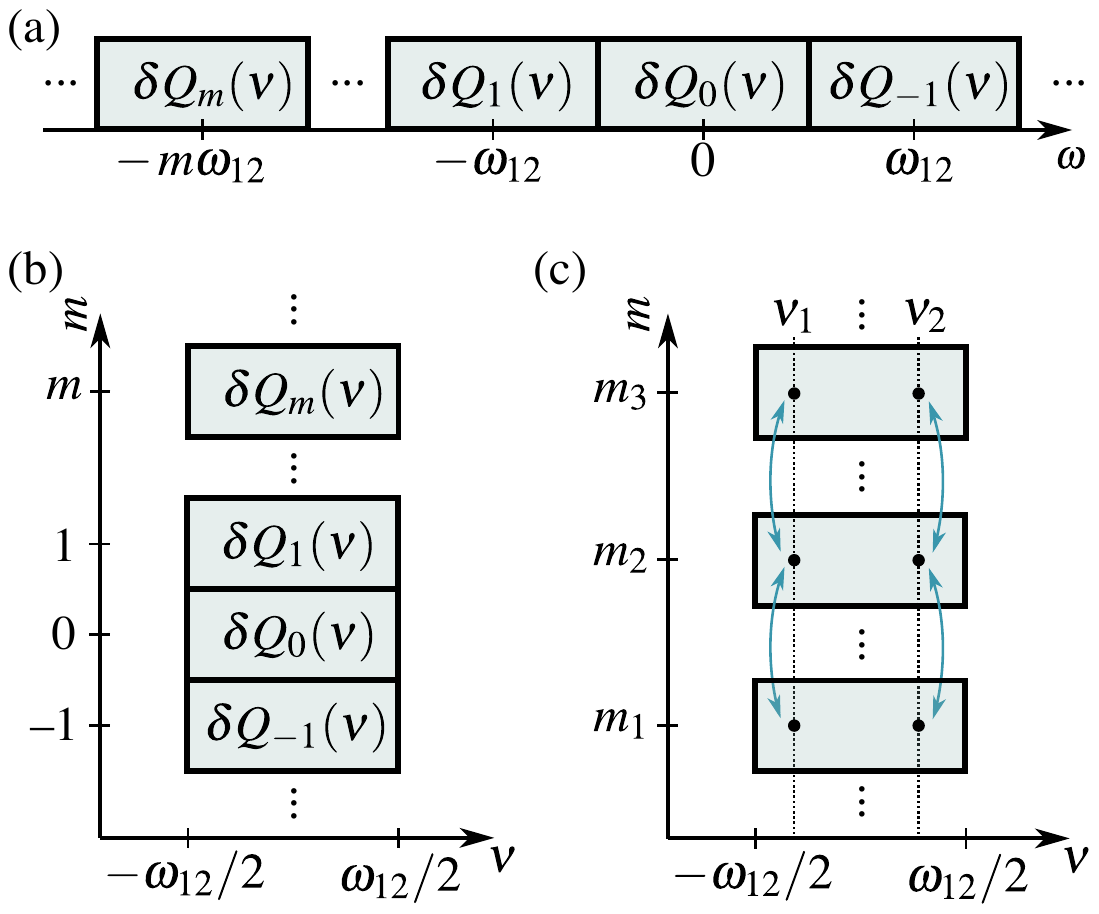}
  \caption{(a) Partitioning of the continuous frequency axis for the fluctuation operators $\delta Q(\omega)$ into zones of width $\omega_{12}$. The continuous frequency $\nu$ is defined to be in the interval $\nu\in[-\omega_{12}/2,\omega_{12}/2]$, and the Fourier index $m$ indicates the zone on the frequency axis, such that $\delta Q_m(\nu):=\delta Q(\omega-m\omega_{12})$.
  (b) The frequency-axis partitioning can be illustrated as a stacking of the frequency zones, such that the continuous frequency $\nu$ is on the horizontal axis and the discrete Fourier index $m$ is on the vertical axis. (c) In this stacked representation, the dynamics of the Heisenberg-Langevin equation Eq.~\eqref{eq:fluctuation-eom-1} only couples fluctuations that are on the same vertical line, i.e. the equations of motion are diagonal in $\nu$. The couplings of Eq.~\eqref{eq:fluctuation-eom-1} are indicated with blue arrows.}
  \label{fig:fourier-band}
\end{figure}

To handle this challenge, we partition the frequency axis into zones of size $\omega_{12}$ and define the frequency $\nu$ to be in the interval $[-\omega_{12}/2,\omega_{12}/2]$. We then introduce the discrete Fourier-index notation $\delta Q_m(\nu) := \delta Q(\nu-m\omega_{12})$ [see Fig.~\ref{fig:fourier-band}(a)--(b)]. Now the Heisenberg-Langevin equations are diagonal in $\nu$, i.e. $\delta Q_m(\nu)$ is not coupled to $\delta Q_{m'}(\nu')$ with $\nu'\neq\nu$, but only to $\delta Q_{m'}(\nu)$. Instead, the frequency off-diagonal coupling in the fluctuation equations appears as a coupling between different Fourier indices [see Fig.~\ref{fig:fourier-band}(c)]. At the same time, we suppress the momentum-index 0 on $\delta P$ and $\delta a$ and use only the Fourier index, such that $\delta P^\dagger_{\zeta,m}(\nu)$ and $\delta a^\dagger_{\zeta,m}(\nu)$ implicitly refer to the zero-momentum exciton and cavity fluctuation operators.

By solving the frequency-space equations for $\delta \mathcal{B}^{\zeta\zeta'}_{\mu,\pm},\;\delta\mathcal{C}^{\zeta\zeta'}_\bq$ and $\delta\mathcal{C}^{\zeta\zeta'}_\bq$ formally (see Appendix~\ref{sec:formal-solutions}), they are eliminated and replaced by a renormalization of the equation for $\delta P^\dagger_{\zeta,m}(\omega)$. For a compact notation, we introduce the combined fluctuation vector
\begin{align}
\label{eq:psi-definition}
\delta \psi_{\zeta,m}(\nu) = [\delta a^\dagger_{\zeta,m}(\nu), \delta P^\dagger_{\zeta,m}(\nu),\delta a_{\zeta,m}(\nu), \delta P_{\zeta,m}(\nu)]^{\rm T}.
\end{align}
The frequency-space Heisenberg-Langevin equation for $\delta \psi$ reads
\begin{align}
\label{eq:psi-H-L-1}
  \sum_{\zeta' m'} [G^{-1}(\nu)]^{mm'}_{\zeta\zeta'} \delta \psi_{\zeta',m'}(\nu)
  = T^{mm'}_{\zeta\zeta'}(\nu)\delta\psi^{\rm in}_{\zeta',m'}(\nu),
\end{align}
where the inverse Green's function is given by
% \begin{widetext}
% \begin{align}
% \label{eq:inverse-GF}
% \begin{split}
%   [G^{-1}(\nu)]^{mm'}_{\zeta\zeta'}
%   &=
%   \delta_{mm'}\delta_{\zeta\zeta'}
%   \mqty[-\hbar\nu_m&0&0&0\\0&-\hbar\nu_m&0&0\\0&0&\hbar\nu_m&0\\0&0&0&\hbar\nu_m  ]
%   \\
%   &- \mqty[\delta_{mm'}\delta_{\zeta\zeta'}(\tilde{E}^{\rm p}_0-\omega_{\rm r}) & \delta_{mm'}\delta_{\zeta\zeta'}\Omega_0 & 0 & 0
%     \\
%       \hat{\Omega}^{mm'}_{\zeta\zeta',0}(\nu) & \delta_{mm'}\delta_{\zeta\zeta'}(\tilde{E}^{\rm x}_0-\omega_{\rm r}) + \Sigma^{mm'}_{\zeta\zeta'}(\nu) & 0 & \Delta^{m'-m}_{\zeta\zeta'}
%     \\
%       0&0&\delta_{mm'}\delta_{\zeta\zeta'}(\tilde{E}^{\rm p*}_0-\omega_{\rm r}) & \Omega_0\delta_{mm'}\delta_{\zeta\zeta'}
%     \\
%       0& \Delta^{m-m'*}_{\zeta\zeta'}&\hat{\Omega}^{-m,-m'}_{\zeta\zeta',0}(-\nu)& \delta_{mm'}\delta_{\zeta\zeta'}(\tilde{E}^{\rm x*}_0-\omega_{\rm r})
%       + \Sigma^{-m,-m'}_{\zeta\zeta'}(-\nu)],
% \end{split}
% \end{align}
% \end{widetext}
\begin{align}
\label{eq:inverse-GF}
\begin{split}
  &[G^{-1}(\nu)]^{mm'}_{\zeta\zeta'}
  = \delta_{mm'}\delta_{\zeta\zeta'}\mathbb{I}_{4}\omega_{\rm r}
  \\
  &+\delta_{mm'}\delta_{\zeta\zeta'}\!\!
  \mqty[
  -\hbar\nu_m - \tilde{E}^{\rm p}_0&0&0&0
  \\0&-\hbar\nu_m - \tilde{E}^{\rm x}_0&0&0
  \\0&0&\hbar\nu_m - \tilde{E}^{\rm p*}_0&0
  \\0&0&0&\hbar\nu_m - \tilde{E}^{\rm x*}_0]
  \\
  &- \mqty[0 & \delta_{mm'}\delta_{\zeta\zeta'}\Omega_0 & 0 & 0
    \\
      \hat{\Omega}^{mm'}_{\zeta\zeta',0}(\nu) & \Sigma^{mm'}_{\zeta\zeta'}(\nu) & 0 & \Delta^{m'-m}_{\zeta\zeta'}
    \\
      0&0& & \Omega_0\delta_{mm'}\delta_{\zeta\zeta'}
    \\
      0& \Delta^{m-m'*}_{\zeta\zeta'}&\hat{\Omega}^{-m,-m'}_{\zeta\zeta',0}(-\nu)&
      \Sigma^{-m,-m'}_{\zeta\zeta'}(-\nu)],
\end{split}
\end{align}
where $\mathbb{I}_4$ is the $4\times 4$ identity matrix, $\nu_m:=\nu-m\omega_{12}$, $\Sigma^{mm'}_{\zeta\zeta'}(\nu)$ is the exciton self-energy, and $\hat{\Omega}_{\zeta\zeta',0}^{mm'}$ is a renormalised exciton-photon coupling strength. The latter two stem from renormalisations due to the formal elimination of the multiparticle fluctuations and are given in Appendix~\ref{sec:formal-solutions}. These renormalisations give small quantitative corrections to the spontaneous four-wave mixing spectra. Most importantly, $\Delta^{m}_{\zeta\zeta'}$ is the parametric gain connecting different fourier components $m_1$ and $m_1+m$, which represents the core mechanism behind spontaneous four-wave mixing. It is given by
\begin{align}
\label{eq:Delta-matrix}
  \begin{split}
    \Delta_{\zeta\zeta'}^m &= \sum_{n}\delta_{\zeta,\zeta'}W^0 \ev*{P_{\zeta,0}^\dagger}_{n}
    \ev*{P_{\zeta,0}^\dagger}_{m-n}
    + \sum_{\mu\pm} W_{\mu}^\pm \mathcal{B}^{\zeta\zeta'}_{\mu\pm,m}
    \\
    &-\delta_{\zeta,\zeta'}\sum_\bq \tilde{\Omega}_\bq\big[ \mathcal{C}^{\zeta\zeta'}_{\bq,m}
    +
    \delta_{\bq,0}\sum_{n}\ev*{a_{\zeta,0}^\dagger}_{n}\ev*{P_{\zeta,0}^\dagger}_{m-n} \big].
  \end{split}
\end{align}
The parametric gain describes a coherent pairwise driving of the excitons, and is a purely a result of the nonlinearities in the system, specifically the Coulomb interaction and the Pauli-blocking from the fermionic substructure of excitons. Moreover, without parametric gain, the generated field would have the same spectral and coherence properties as the classical input field. The parametric gain contains terms from the instantaneous Coulomb interaction between excitons ($W^0$), biexcitonic correlations ($W^\pm_\mu$) from the bound biexciton ($\mu=\mathrm{b}$) and two-exciton continuum ($\mu\neq\mathrm{b}$), and Pauli-blocking $\tilde{\Omega}_\bq$. All matrix elements and coefficients are given in Appendices~\ref{sec:matrix-elements} and~\ref{sec:formal-solutions}.

%In the limit $\Delta^{m}_{\zeta\zeta'}\rightarrow 0$, no spontaneous four-wave mixing occurs.
The matrix $T^{mm'}_{\zeta\zeta'}(\nu)$ describes the coupling to the input field, $\delta \psi^{\rm in}_{\zeta,m}(\nu)=[\delta a^{\rm in\dagger}_{\zeta,m}(\nu), \delta P^{\rm in\dagger}_{\zeta,m}(\nu),\delta a^{\rm in}_{\zeta,m}(\nu), \delta P^{\rm in}_{\zeta,m}(\nu)]$. The contribution to $T^{mm'}_{\zeta\zeta'}(\nu)$ in zeroth order of the input field is given by $T^{mm'}_{\zeta\zeta'}(\nu) = i\hbar\delta_{mm'}\delta_{\zeta\zeta'}\:{\rm diag}[\sqrt{2\gamma^{\rm p}},\sqrt{2\gamma^{\rm x}},-\sqrt{2\gamma^{\rm p}},-\sqrt{2\gamma^{\rm p}}]$. Due to elimination of the multiparticle fluctuations, additional second-order contributions are also present, which are given in Appendix~\ref{sec:formal-solutions}.

The formal solution of Eq.~\eqref{eq:psi-H-L-1} is given by
\begin{align}
  \delta\psi_{m,\zeta}(\nu) = \sum_{m'\zeta'}\mathcal{G}^{mm'}_{\zeta\zeta'}(\nu)\delta\psi^{\rm in}_{\zeta'm'}(\nu),
\end{align}
where $\mathcal{G}^{mm'}_{\zeta\zeta'}(\nu) = \sum_{m''\zeta''}G^{mm''}_{\zeta\zeta''}(\nu)T^{m''m'}_{\zeta''\zeta'}(\nu)$.
We note that in practise, the renormalizations from formal elimination of the multiparticle fluctuations $\delta \mathcal{B},\;\delta \mathcal{C}$ and $\delta \mathcal{D}$ influence the numerical calculations presented in this paper only with small quantitative corrections. This means that in many cases, one can neglect $\delta \mathcal{B}$ and $\delta \mathcal{C}$ in Eq.~\eqref{eq:fluctuation-eom-1}, leading to the simplifications
\begin{align}
\label{eq:DCTA-simplifications}
  \begin{split}
    \Sigma^{mm'}_{\zeta\zeta'}(\nu)&=0, \\
    \hat{\Omega}^{mm'}_{\zeta\zeta'0}(\nu) &= \Omega_0\delta_{\zeta\zeta'}\delta_{mm'}\\
    T^{mm'}_{\zeta\zeta'}(\nu) &= i\hbar\delta_{mm'}\delta_{\zeta\zeta'}\:{\rm diag}[\sqrt{2\gamma^{\rm p}},\sqrt{2\gamma^{\rm x}},-\sqrt{2\gamma^{\rm p}},-\sqrt{2\gamma^{\rm p}}].
  \end{split}
\end{align}
This is demonstrated explicitly in Appendix~\ref{sec:appendix-multiparticle-effect}, where we compare the full calculation of fluctuation spectra with the calculation under the simplifications in Eq.~\eqref{eq:DCTA-simplifications}. All calculations presented in the main text have been performed using the full contributions of Eq.~\eqref{eq:inverse-GF}.
%Due to the negligible effect of the multiparticle fluctuation renormalizations, we use the simplified theory for the calculations presented in the main text.
However, the negligible influence of multiparticle fluctuations supports the linearization of the Heisenberg-Langevin equations, since products such as $\delta\mathcal{C}\delta P$ can be expected to have an even smaller effect than the terms linear in $\delta \mathcal{C}$. This fact does not mean that many-body correlation effects are unimportant: indeed, the dominating contribution to the parametric gain stems from the bound biexciton, $\mathcal{B}_{\rm b,-}^{KK'}$.

\subsection{Emission spectrum and squeezing}
\label{sec:emis-spec-squeez}
The detected field is the output from the cavity perpendicular to the surface, i.e. at zero in-plane momentum $\bk=0$. We take the field to be polarisation-filtered before detection and denote the polarisation-projected cavity  operator by $a = \bm{\lambda}_{\rm out}^\mathrm{T}\mathbf{a}_0$~\cite{gardiner2004quantum}, where $\mathbf{a}_0 = (a_{K,0}, a_{K',0})^{\rm T}$ and $\bm{\lambda}_{\rm out}$ is the polarisation of the detection channel in the circular basis. The corresponding fluctuation operator is $\delta a(t) = \lim_{t\rightarrow\infty}[a(t)-\ev*{a(t)}]$.

\subsubsection{Spectral correlation functions}
The emission properties of the generated field from the cavity are characterized by two spectral correlation functions
\begin{align}
\begin{split}
\label{eq:spectral-correlation-func-S1-1}
\ev*{\delta a_m^\dagger(\nu)\delta a_{m'}(\nu')} &=
\sum_{\zeta\zeta'}\sum_{\zeta_1\zeta_2}\sum_{m_1m_2}\sum_{j_1j_2}
\lambda_{\rm out}^\zeta\lambda_{\rm out}^{\zeta'}[\mathcal{G}_{\zeta\zeta_1}^{mm_1}(\nu)]_{1,j_1}
\\ &\times[\mathcal{G}_{\zeta'\zeta_2}^{m'm_2}(\nu')]_{3,j_2}
\ev*{\delta\psi^{{\rm in},j_1}_{\zeta_1 m_1}(\nu)\delta\psi^{{\rm in},j_2}_{\zeta_2 m_2}(\nu')}
\end{split}
\end{align}
\begin{align}
\begin{split}
\label{eq:spectral-correlation-func-S2-1}
\ev*{\delta a_m(\nu)\delta a_{m'}(\nu')} &=
\sum_{\zeta\zeta'}\sum_{\zeta_1\zeta_2}\sum_{m_1m_2}\sum_{j_1j_2}
\lambda_{\rm out}^\zeta\lambda_{\rm out}^{\zeta'}[\mathcal{G}_{\zeta\zeta_1}^{mm_1}(\nu)]_{3,j_1}
\\ &\times[\mathcal{G}_{\zeta'\zeta_2}^{m'm_2}(\nu')]_{3,j_2}
\ev*{\delta\psi^{{\rm in},j_1}_{\zeta_1 m_1}(\nu)\delta\psi^{{\rm in},j_2}_{\zeta_2 m_2}(\nu')}
\end{split}
\end{align}
where $j\in[1,4]$ is the vector index of $\delta\psi$ as defined in Eq.~\eqref{eq:psi-definition}. As shown in Sections~\ref{sec:emis-spec} and \ref{sec:homodyne}, the emission spectrum is described through the correlation function in Eq.~\eqref{eq:spectral-correlation-func-S1-1}, whereas the squeezing spectrum is described through both correlation functions in Eqs.~\eqref{eq:spectral-correlation-func-S1-1} and~\eqref{eq:spectral-correlation-func-S2-1}.

The spectral correlation of the input field is derived from the temporal correlations from Sec.~\ref{sec:fluctuations} as
\begin{align}
  \ev*{\delta\psi^{{\rm in},j}_{\zeta m}(\nu)\delta\psi^{{\rm in},j'}_{\zeta' m'}(\nu')} = 2\pi\delta_{\zeta\zeta'}\delta_{m,-m'}\eta_{jj'}\delta(\nu+\nu'),
\end{align}
where $\eta_{jj'} = \delta_{j',1}\delta_{j,3} + \delta_{j',2}\delta_{j,4}$. Although no correlations are present between photon and exciton input fields, such correlations will generally be nonzero in the internal field $\delta \psi$. With this, we obtain
\begin{align}
  \ev*{\delta a_{-m}^\dagger(-\nu)\delta a_{m'}(\nu')} &= 2\pi S_1^{mm'}(\nu)\delta(\nu-\nu')
  \\
  \ev*{\delta a_{-m}(-\nu)\delta a_{m'}(\nu')} &= 2\pi S_2^{mm'}(\nu)\delta(\nu-\nu'),
\end{align}
where
\begin{align}
\label{eq:S12-spectral-correlation-functions}
\begin{split}
  S_1^{mm'}(\nu) &= \sum_{\zeta\zeta'}\sum_{\zeta_1 m_1}\sum_{j_1j_2}
  \lambda_{\rm out}^\zeta\lambda_{\rm out}^{\zeta'}[\mathcal{G}_{\zeta\zeta_1}^{-m,-m_1}(-\nu)]_{1,j_1}
  \\ &\hspace{2.5cm}\times[\mathcal{G}_{\zeta'\zeta_1}^{m',m_1}(\nu)]_{3,j_2}\eta_{j_1 j_2}
  \\
  S_2^{mm'}(\nu) &= \sum_{\zeta\zeta'}\sum_{\zeta_1 m_1}\sum_{j_1j_2}
  \lambda_{\rm out}^\zeta\lambda_{\rm out}^{\zeta'}[\mathcal{G}_{\zeta\zeta_1}^{-m,-m_1}(-\nu)]_{3,j_1}
  \\ &\hspace{2.5cm}\times[\mathcal{G}_{\zeta'\zeta_1}^{m',m_1}(\nu)]_{3,j_2}\eta_{j_1 j_2}.
\end{split}
\end{align}
For convenience, we shall use the shorthand notation
\begin{align}
\label{eq:Si-w-continuous-frequency}
S_i(\nu-m\omega_{12}) := S^{mm}_i(\nu),
\end{align}
in order to label the fluctuation spectra on the continuous frequency axis rather than in the Fourier-zone partitioning. This relation can alternatively be cast as
$$S_i(\omega) := \sum_m\int\dd{\nu} S_i^{mm}(\nu)\delta(\omega-(\nu-m\omega_{12}))$$.

\subsubsection{Emission spectrum}
\label{sec:emis-spec}
The emission spectrum of the outcoupled field $S_{\rm tot}(\omega) = S_{\rm coh}(\omega) + S_{\rm FWM}(\omega)$ contains a coherent contribution $S_{\rm coh}=S_{\rm coh,1}\delta(\omega-\omega_1) + S_{\rm coh,2}\delta(\omega-\omega_2)$, which shares the spectral distribution and temporal coherence of the driving field~\cite{steck2007quantum,gardiner2004quantum}, and a spontaneously generated field $S_{\rm FWM}$, which is created by spontaneous four-wave mixing. The former is of minor interest in the present investigation, because it simply shares the quantum statistics with the classical input field. The primary emission spectrum of interest is the spontaneous four-wave mixing spectrum, which is given by
\begin{align}
  S_{\rm FWM}(\omega) = 2\gamma^{\rm p}S_1(\omega),
\end{align}
where $S_1(\omega)$ is defined in Eq.~\eqref{eq:Si-w-continuous-frequency}.

\subsubsection{Homodyne noise spectrum and quadrature squeezing}
\label{sec:homodyne}

\begin{figure}
\includegraphics[width=\columnwidth]{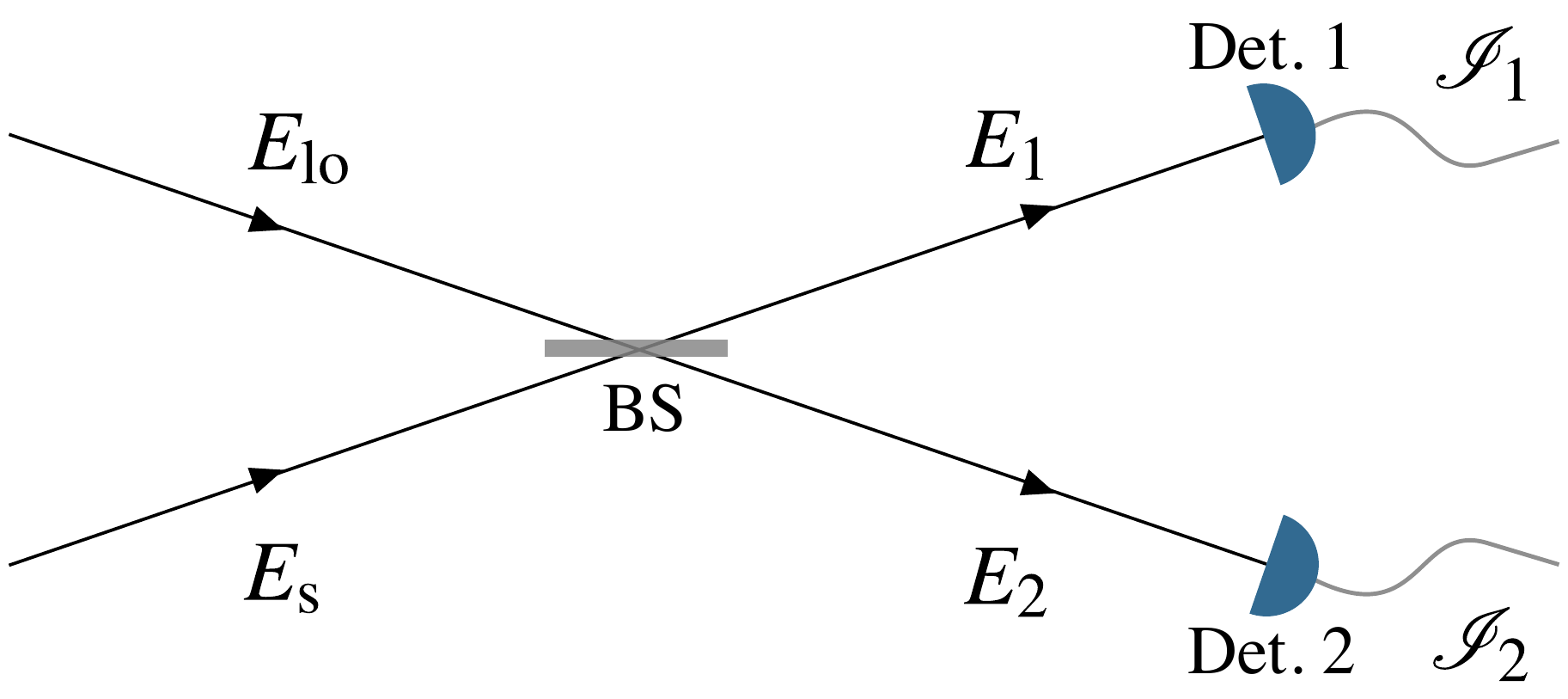}
  \caption{Homodyne detection. The source field $E_{\rm s}$ is mixed with a strong, coherent local oscillator field $E_{\rm lo}$ on a 50/50 beamsplitter (BS). The outgoing field modes $E_1$ and $E_2$ are detected on detectors 1 and 2, which produces the photocurrents $\mathscr{I}_1$ and $\mathscr{I}_2$. The homodyne signal is the current difference $\mathscr{I}_1-\mathscr{I}_2$.}
  \label{fig:homodyne}
\end{figure}

Quadrature squeezing is described through homodyne detection of the output field from the cavity (see Fig.~\ref{fig:homodyne}). While the quadrature squeezing of the internal cavity mode can be evaluated simply as the variance of the quadrature operator $\delta X(\theta) := e^{i\theta}\delta a^\dagger + e^{-i\theta}\delta a$ with respect to the quadrature phase $\theta$, the outcoupled and thus measurable field quadrature fluctuations are more complicated~\cite{gardiner2004quantum,carmichael1987spectrum}. In homodyne detection, the source field from the cavity $E_{\rm s}(t) = \sqrt{2\gamma^{\rm p}} a(t)$ is mixed with a strong local oscillator $E_{\rm lo}(t)$ on a beamsplitter, such that the fields leaving the beamsplitter $E_1$ and $E_2$ are given by
\begin{align}
\label{eq:homodyne-beamsplitter}
  \mqty[E_1(t) \\ E_2(t)] = \frac{1}{\sqrt{2}}\mqty[1&i\\i&1]\mqty[E_{\rm s}(t) \\ E_{\rm lo}(t)].
\end{align}
The local oscillator has a frequency equal to the rotating frame frequency $\frac{1}{2}(\omega_1+\omega_2)$. Thus, when working in the rotating frame, the local oscillator expectation values are time-independent: $\ev*{E_{\rm lo}(t)}=e^{i\varphi}\sqrt{F_{\rm lo}}$, where $F_{\rm lo}$ and $\varphi$ are the photon flux and reference phase of the local oscillator, respectively, and the expectation value is evaluated in the rotating frame. Taking the local oscillator to be in a coherent state, we have $\ev*{E_{\rm lo}^\dagger(t) E_{\rm lo}(t)} = F_{\rm lo}$. In the homodyne measurement process, the two output fields $E_1$ and $E_2$ are measured on separate photodetectors (which are labeled with corresponding indices 1 and 2) and the difference of these photocurrents are recorded as the signal.

The statistical properties of this photocurrent difference are derived using normal-ordered detection theory~\cite{glauber1963quantum,kelley1964theory,paul1988shot,collett1987quantum,carmichael1987spectrum}.
Here, we extend the derivation of Ref.~\onlinecite{carmichael1987spectrum} to the case of balanced homodyne detection and to bichromatic driving, where no time-independent stationary source field exists. The detection model is based on the assumption that a single photoelectric detection event produces a current pulse of duration $\tau_{\rm d}$ and amplitude $ge/\tau_{\rm d}$, where $e$ and $g$ denote the electronic charge and photodetector gain, respectively. The photocurrents of detector $\mu=1,2$ is then given by~\cite{kelley1964theory,carmichael1987spectrum}
\begin{align}
  \label{eq:19}
  \mathscr{I}_\mu(t)=\frac{ge}{\tau_{\rm d}}n_\mu,
\end{align}
where $n_{\mu}$ is a classical stochastic variable representing the number of overlapping pulses in the detection electronics of detector $\mu$, i.e. the number of pulses initiated in the time interval $t-\tau_{\rm d}$ to $t$. For practical purposes, we shall work with a scaled current, $I_\mu(t):=\mathscr{I}_\mu(t)/(ge)$, which has units of inverse time and represents the flux of detected photons.

The joint probability of detecting $n_{1}$ photons in detector 1 and $n_{2}$ in detector 2 within the time interval $[t-\tau_{\rm d},t]$ is given by~\cite{kelley1964theory,carmichael1987spectrum}
\begin{align}
  \label{eq:p1-distribution}
\begin{split}
  p^{(1)}(n_1, &t-\tau_{\rm d}, t; n_2, t-\tau_{\rm d}, t) \\ &=\ev{:\prod_{\mu=1,2}\frac{ [\mathscr{G}_\mu(t-\tau_{\rm d},t)]^{n_{\mu}}}{n_{\mu}!}e^{-\mathscr{G}_\mu(t-\tau_{\rm d},t)}:},
\end{split}
\end{align}
where the generator $\mathscr{G}$ of the distribution is given by
\begin{align}
  \mathscr{G}_\mu(t_1,t_2)=\eta \int_{t_1}^{t_2}\dd{t} E_\mu^\dagger(t)E_\mu(t),
\end{align}
and $\eta$ is the detection efficiency. The symbol $::$ denotes normal- and time-ordering at the level of the electric field operators ($E^\dagger_\mu$ arranged to the left of $E_\mu$ and time arguments increasing to the right (left) in products of $E_\mu^\dagger$ ($E_\mu$)). The probability distribution Eq.~\eqref{eq:p1-distribution} is essentially Mandel's counting formula, as presented in e.g. Ref.~\cite{gardiner2004quantum}. From $p^{(1)}$, the mean photocurrent is derived as~\cite{carmichael1987spectrum}
\begin{align}
\label{eq:mean-photocurrent}
\begin{split}
\overline{I_\mu(t)} &=\frac{1}{\tau_{\rm d}} \ev{\mathscr{G}_\mu(t-\tau_{\rm d},t)},
\\
&=\frac{\eta}{\tau_{\rm d}}\int_{t-\tau_{\rm d}}^t\dd{t'} \ev*{E_\mu^\dagger(t') E_\mu(t')}
\end{split}
\end{align}
where we note that we use overlines to denote the mean value of the classical stochastic variable $I_\mu$ and angular brackets to denote quantum-mechanical expectation values.

Similarly, the two-time probability of detecting $n$ counts in detector $\mu$ in the time interval $[t-\tau_{\rm d},t]$ and $m$ counts in detector $\mu'$ in the time interval $[t+\tau-\tau_{\rm d},t+\tau]$ is given by~\cite{carmichael1987spectrum}
\begin{align}
  \label{eq:p2-distribution}
\begin{split}
  &p^{(2)}_{\mu\mu'}(n,t-\tau_{\rm d},t;m,t+\tau-\tau_{\rm d},t+\tau)
\\ &= \Bigg\langle :\frac{\mathscr{G}_\mu(t-\tau_{\rm d},t)^n}{n!}e^{-\mathscr{G}_\mu(t-\tau_{\rm d},t)}\\ &\hspace{1cm}\times \frac{\mathscr{G}_{\mu'}(t+\tau-\tau_{\rm d},t+\tau)^m}{m!}e^{-\mathscr{G}_{\mu'}(t+\tau-\tau_{\rm d},t+\tau)}:\Bigg\rangle.
\end{split}
\end{align}
From this probability, the detector current correlation function is found to be (for $\tau>0$)
\begin{align}
\label{eq:photocurrent-correlation-1}
\begin{split}
  \overline{I_\mu(t)I_{\mu'}(t+\tau)}&= \\ \qty(\frac{1}{\tau_{\rm d}})^2&\Big[\ev{:\mathscr{G}_\mu(t-\tau_{\rm d},t)\mathscr{G}_{\mu'}(t+\tau-\tau_{\rm d},t+\tau):} \\
  &+ \delta_{\mu\mu'}\Theta(\tau_{\rm d}-\tau)\ev{\mathscr{G}_\mu(t+\tau-\tau_{\rm d},t)}\Big],
\end{split}
\end{align}
where $\Theta$ is the Heaviside function.

The measured spectral noise function $N(\omega)$ of the homodyne signal $I_-:=I_1-I_2$ is then the Fourier transformation of the photocurrent fluctuation correlation function, averaged over one period of the signal oscillation $T=2\pi/\abs{\omega_{12}}=4\pi/\abs{\omega_2-\omega_1}$:
\begin{align}
\label{eq:Nw-definition}
\begin{split}
N(\omega,\theta) = &\lim_{t_0\rightarrow\infty}\frac{1}{T}\int_{t_0}^{t_0+T}\dd{t}\int_0^\infty\dd{\tau} \cos(\omega\tau) \\ &\hspace{2cm}\times\qty[\overline{I_-(t)I_-(t+\tau)}- \overline{I_-(t)}\;\;\overline{I_-(t+\tau)}].
\end{split}
\end{align}

From Eqs.~\eqref{eq:homodyne-beamsplitter}, \eqref{eq:mean-photocurrent} and \eqref{eq:photocurrent-correlation-1}, the following expression for the photocurrent fluctuation correlation function is derived (see Appendix~\ref{sec:appendix-homodyne})
\begin{align}
\label{eq:photocurrent-correlation-3}
\begin{split}
  &\overline{I_-(t)I_-(t+\tau)} -
  \overline{I_-(t)}\;\;\overline{I_-(t+\tau)}
\\
  &= N_0(\tau) + \frac{\eta^2F_{\rm lo}}{\tau_{\rm d}^2}
  \int_{t-\tau_{\rm d}}^t\!\!\!\!\!\!\dd{t'}
  \int_{t-\tau_{\rm d}}^{t}\!\!\!\!\!\!\dd{t''}
  \ev*{:\delta X_{\rm s}(\theta,t')\delta X_{\rm s}(\theta,t''+\tau):},
\end{split}
\end{align}
where $X_{\rm s}(\theta,t) = e^{i\theta} E_{\rm s}^\dagger(t) + e^{-i\theta}E_{\rm s}(t)$ is the source-field quadrature operator with angle $\theta=\varphi+\pi/2$ and its fluctuation operator is $\delta X_{\rm s}(\theta,t) := X_{\rm s}(\theta,t)-\ev*{X_{\rm s}(\theta,t)}$. The measured quadrature angle $\theta$ stems from the local oscillator phase $\varphi$ and a phase displacement $\pi/2$ from the beamsplitter as seen in Eq.~\eqref{eq:homodyne-beamsplitter}. The first term in Eq.~\eqref{eq:photocurrent-correlation-3} is the shot noise correlation function, which in the limit where the local oscillator is much stronger than the source field is given by
\begin{align}
\label{eq:shot-noise}
  N_0(\tau) = \frac{\eta F_{\rm lo}}{\tau_{\rm d}^2}(\tau_{\rm d}-\tau)\Theta(\tau_{\rm d}-\tau).
\end{align}
In order to characterize the intrinsic noise properties of the source field and to disentangle these properties from the detector properties, we take the limit of infinite detection bandwidth ($\tau_{\rm d}\rightarrow 0$) and unity detection efficiency ($\eta=1$). In this limit, inserting Eq.~\eqref{eq:photocurrent-correlation-3} into Eq.~\eqref{eq:Nw-definition} yields the noise spectrum relative to the shot-noise level (see Appendix~\ref{sec:appendix-homodyne})
\begin{align}
  \frac{N(\omega)}{N_0(\omega)} = 1+\Lambda(\omega,\theta),
\end{align}
where $N_0(\omega)=\int_0^\infty\dd{\tau}\cos(\omega\tau)N_0(\tau)$ is the shot noise spectrum and
\begin{align}
\label{eq:Nw-2}
  \Lambda(\omega,\theta) = {\rm Re}[\Lambda_1(\omega) + e^{-2i\theta}\Lambda_2(\omega)],
\end{align}
where $\Lambda_i(\omega) = 4\int_0^\infty\dd{\tau}\cos(\omega\tau)C_i(\tau)$
and
$C_i(\tau) = \frac{1}{2\pi}
\int_{-\infty}^\infty \dd{\omega}
e^{i\omega\tau}
2\gamma^{\rm p}S_i(\omega)$. All details of the derivation are given in Appendix~\ref{sec:appendix-homodyne}.

 The optimal quadrature angle, where the noise is minimized, is given by $e^{-2i\theta}=-\Lambda_2(\omega)/\abs{\Lambda_2(\omega)}$. At this angle, the spectrum of squeezing is
\begin{align}
  \Lambda(\omega) = {\rm Re}[\Lambda_1(\omega)] - \abs{\Lambda_2(\omega)},
\end{align}
where the homodyne phase $\theta$ has been suppressed in $\Lambda(\omega,\theta)$ for notational brevity. In the analysis of squeezing, we shall use the normalized optimal-angle spectrum $1+\Lambda(\omega)$ to characterize the amount of squeezing in the generated light.
This spectrum is normalized such that a value of $1+\Lambda(\omega)=1$ corresponds to the shot noise level, i.e. no squeezing at all, whereas a value of $1+\Lambda(\omega)=0$ corresponds to complete elimination of the shot noise and thus perfect squeezing.

We note that the spontaneous four-wave mixing spectrum $S_{\rm FWM}(\omega)$ and the squeezing spectrum $\Lambda(\omega)$ are of fundamentally different nature and are measured with very different techniques. The spontaneous four-wave mixing spectrum $S_{\rm FWM}(\omega)$ can be detected by filtering out the bichromatic pump from the generated light and sending the filtered signal into an optical spectrum analyser. The frequency argument $\omega$ is connected to the emitted photon energy on the order of 2 eV. In contrast, the squeezing spectrum $\Lambda(\omega)$ is measured through homodyne detection, i.e. by beating the generated signal with a local oscillator at frequency $\omega_{\rm r}:=(\omega_1+\omega_2)/2$. Here, the frequency argument $\omega$ is the beat frequency between the signal and the local oscillator, and is thus much smaller, on the order of a few meV.

\section{Results}
\label{sec:results}

\begin{figure*}
\centering
\includegraphics[width=\textwidth]{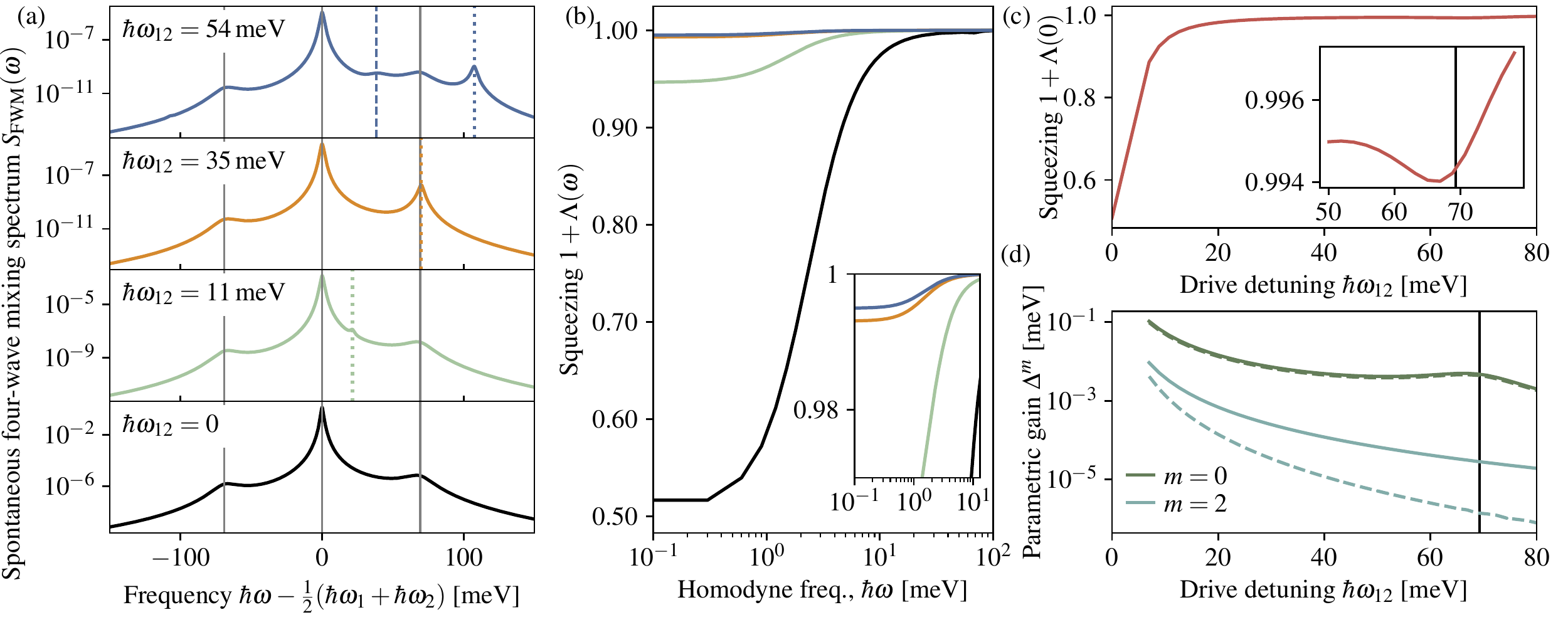}
  \caption{Results for Configuration A. (a) Spontaneous four-wave mixing spectrum for four values of the drive frequency difference $\hbar\omega_{12}:=\frac{1}{2}(\omega_2-\omega_1)$. The vertical lines indicate four-wave mixing resonances as described in the main text. (b) Homodyne squeezing spectra as $1+\Lambda(\omega)$ with colours corresponding to the values of $\hbar\omega_{12}$ in panel (a). The inset shows a zoom-in region at the upper section of the y-axis. (c) Zero-frequency squeezing $1+\Lambda(0)$ as a function of the drive frequency difference $\hbar\omega_{12}$. The inset shows a zoom-in region around the weak resonance at $\hbar\omega_{12}=E^+_0-E^-_0$ (indicated with vertical line). (d) Fourier components $m=0$ and $m=2$ of parametric gain $\Delta^m$, shown as the sum of the norms of spin-diagonal and spin off-diagonal matrix elements. The full lines shows the total parametric gain, and the dashed lines show the contributions from the bound biexciton alone. The vertical line indicates the weak resonance at $\hbar\omega_{12}=E^+_0-E^-_0$.}
  \label{fig:config_A}
\end{figure*}

By numerically solving the equations of motion Eq.~\eqref{eq:eom} explicitly to reach the periodic steady state \rev{(see Appendix~\ref{sec:appendix-numerical-calculations})}, the corresponding Fourier components are calculated from the resulting time series over a steady-state period from Eq.~\eqref{eq:discrete-fourier-definition}. The steady-state Fourier components are then used to construct the fluctuation Green's function from Eq.~\eqref{eq:inverse-GF}, and calculate the spontaneous four-wave mixing spectrum $S_{\rm FWM}(\omega)$ and the squeezing spectrum $\Lambda(\omega)$ as described in Sec.~\ref{sec:emis-spec-squeez}. In this section, we present these results for Configurations A and B, respectively, as shown in Fig~\ref{fig:1}(c)-(d). The numerical calculations become increasingly challenging as $\hbar\omega_{12}:=(\omega_2-\omega_1)/2$ approaches 0, because the oscillation period becomes longer and longer. However, in the case where $\hbar\omega_{12}=0$, the spectra $S_{\rm FWM}(\omega)$ and $\Lambda(\omega)$ can be calculated using the simpler calculation methods for a monochromatic driving field described in Ref.~\onlinecite{denning2022efficient}. For completeness, we include the results for $\hbar\omega_{12}=0$ in this way. The calculations are performed for atomically thin $\mathrm{MoS_2}$ encapsulated by hexagonal BN on both sides. The coupling strength of the cavity is taken to be $\Omega_0 = 20\:{\rm meV}$ and the cavity outcoupling rate is taken to be $\hbar\gamma^{\rm p} = 9\;{\rm meV}$, consistent with fabricated devices~\cite{liu2015strong,anton2021bosonic}. The temperature is taken to be 30 K, leading to a phonon-induced exciton dephasing of $\hbar\gamma^{\rm x}=0.8\;{\rm meV}$ (see Appendix~\ref{sec:appendix-numerical-calculations}). The pump polarization is taken to be linear, with the two drives having the same linear polarization, and the pump power is 5 mW with a pump spot size of 9 $\mathrm{\; \mu m^2}$ . The detected polarization is taken to be co-linear with the drive, and we note that very similar results are seen for cross-polarized detection. Further details about the numerical calculations and all parameters are given in Appendix~\ref{sec:appendix-numerical-calculations}.

\subsection{Configuration A}
\label{sec:config-A}

Configuration A is defined by setting the cavity frequency such that the lower polariton energy $E^-_0$ matches half the bound biexciton energy $\frac{1}{2}E^{\rm xx}_{\rm b,-}$, and setting the mean value of the driving fields to $\frac{1}{2}(\hbar\omega_1+\hbar\omega_2)=E^-_0=\frac{1}{2}E^{\rm xx}_{\rm b,-}$ [cf. Fig.~\ref{fig:1}(c)]. When the drive energy difference $\hbar\omega_{12}$ is zero, the degenerate driving field resonantly creates lower polaritons, which resonantly scatter into bound biexcitons via the Coulomb interaction. The latter process is also known as a polaritonic Feshbach resonance~\cite{wouters2007resonant,carusotto2010feshbach,takemura2014polaritonic} and gives a strong enhancement of the nonlinear response, which in this case leads to strong spontaneous four-wave mixing and single-mode squeezing. Since the in-scattering particles in this configuration (lower polaritons) are identical, this is a \emph{degenerate} Feshbach resonance.

In Fig.~\ref{fig:config_A}(a), the spontaneous four-wave mixing spectrum is shown for four values of the drive frequency difference $\hbar\omega_{12}$. The spectra feature multiple peaks, which we can identify by considering the resonant four-wave mixing channels. Due to energy conservation the energy of a generated photon pair with frequencies $\omega_3$ and $\omega_4$ must match the sum of two pump photon energies. With two different pump frequencies $\omega_1$ and $\omega_2$, there are three possible combinations of pump photon pairs, meaning that the output photon pair frequencies must fulfill one of the relations
\begin{subequations}
\label{eq:fwm-resonances-1}
\begin{align}
  \omega_3+\omega_4 &= \omega_1+\omega_2 \\
  \omega_3+\omega_4 &= 2\omega_1\\
  \omega_3+\omega_4 &= 2\omega_2.
\end{align}
\end{subequations}
The resonant output channels appear where one of the output photon energies equals a polariton energy. Thus, by setting $\hbar\omega_3=E^\pm_0$, the resonance conditions Eq.~\eqref{eq:fwm-resonances-1} become
\begin{subequations}
\begin{align}
\label{eq:w4-a}
  \omega_4 &= \omega_1+\omega_2-E^\pm_0/\hbar \\
\label{eq:w4-b}
  \omega_4 &= 2\omega_1 - E^\pm_0/\hbar \\
\label{eq:w4-c}
  \omega_4 &= 2\omega_2 - E^\pm_0/\hbar.
\end{align}
\end{subequations}
These four-wave mixing resonance conditions give rise to a total of eight possible output photon frequencies (two values of $\omega_3$ and three values of $\omega_4$ for each of these).
In Fig.~\ref{fig:config_A}(a), the resonances $E_0^\pm$ and $\hbar\omega_1+\hbar\omega_2-E^\pm_0$ are indicated with vertical grey lines. These resonance frequencies are independent of $\hbar\omega_{12}$. The strongest (middle) resonance peak is the lower polariton energy $E^-_0$, which in Configuration A matches the mean drive energy, $\frac{1}{2}(\hbar\omega_1+\hbar\omega_2)$. The corresponding photon pair partner from Eq.~\eqref{eq:w4-a} has the same energy. The upper polariton resonance $E^+_0$ and its partner $\hbar\omega_1+\hbar\omega_2-E^+_0$ are positioned symmetrically around the lower polariton. Since the excitation of the upper polariton is much further from resonance, these peaks appear with a much weaker amplitude than the central lower polariton peak.

In addition, the resonance condition in Eq.~\eqref{eq:w4-c} for the lower polariton, $2\hbar\omega_2-E^-_0$ is also shown with vertical dotted lines. This resonance depends on $\hbar\omega_{12}$ and becomes degenerate with $E^-_0$ in the limit $\hbar\omega_{12}=0$.

For the largest value of $\hbar\omega_{12}$, the resonance in Eq.~\eqref{eq:w4-c} is also shown for the upper polariton, i.e. $2\hbar\omega_2-E^+_0$, with a dashed vertical line. This resonance cannot be seen for the other values of $\hbar\omega_{12}$: when $\hbar\omega_{12}$ is too small, the drive energies $\hbar\omega_1$ and $\hbar\omega_2$ are close to resonance with $E^-_0$, but very far away from the upper polariton $E^+_0$. When $\hbar\omega_{12}$ matches the polariton splitting $E^+_0-E^-_0$, as is approximately the case in the panel with $\hbar\omega_{12}=35\;\mathrm{meV}$, the emission line $\hbar\omega_2-E^+_0$ is degenerate with the dominating lower polariton $E^-_0$. When $\hbar\omega_{12}$ is larger, the drives are far away from resonance with the lower polariton, meaning that the dominating polaritonic features (as indicated with vertical gray lines) are diminished, thereby revealing the much weaker peak at $\hbar\omega_2-E^+_0$.

Fig.~\ref{fig:config_A}(b) shows the homodyne squeezing spectrum as $1+\Lambda(\omega)$. As discussed in Section~\ref{sec:homodyne}, this spectrum is normalized such that a value of $1+\Lambda(\omega)=1$ corresponds to no squeezing at all, whereas a value of $1+\Lambda(\omega)=0$ corresponds perfect squeezing. Since the dominant emission channel in Configuration A is degenerate photon pairs at the lower polariton energy $E^-_0$, the output field is single-mode squeezed, which is observed as a squeezing spectrum that is minimal (providing the strongest squeezing) at zero frequency. The squeezing bandwidth is several meV, which stems from the typical scale of the resonance linewidths of the cavity photons, excitons and biexcitons in the system~\cite{hoff2015integrated}.

\begin{figure*}
\centering
\includegraphics[width=\textwidth]{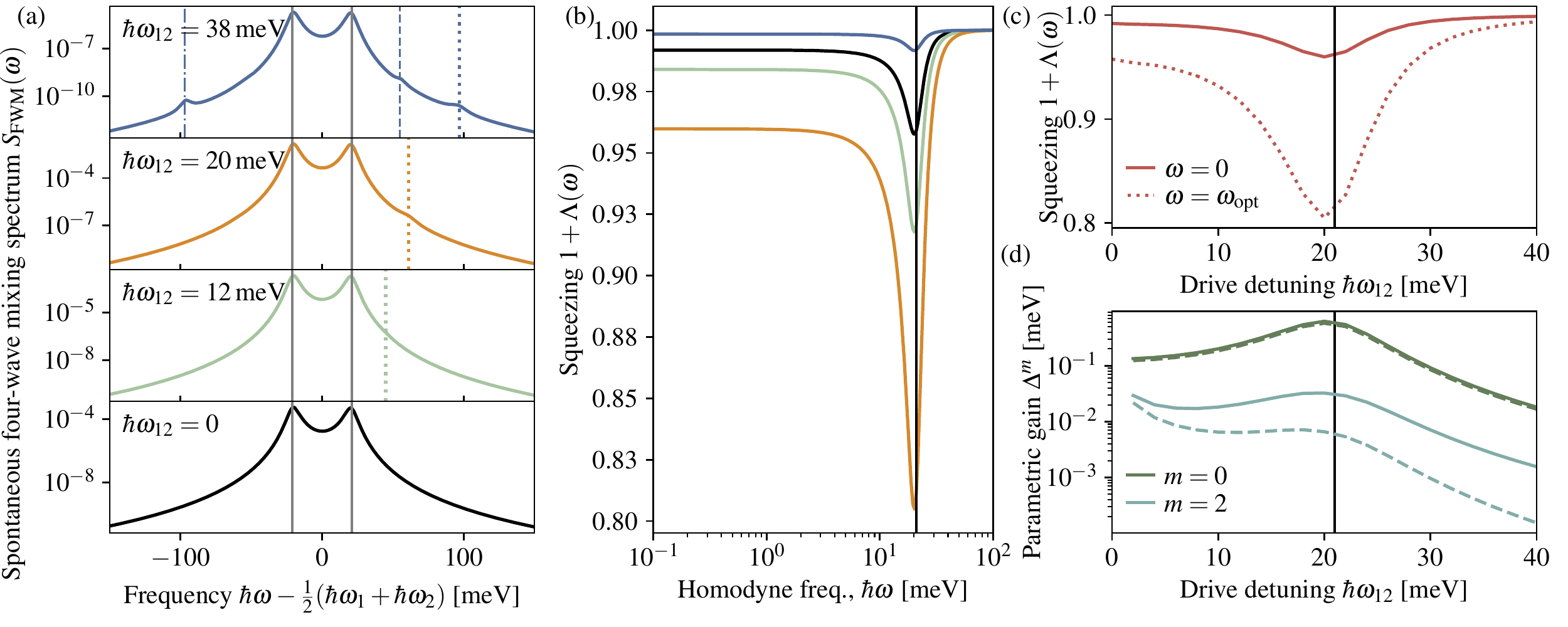}
  \caption{Results for Configuration B. (a) Spontaneous four-wave mixing spectrum for four values of the drive frequency difference $\hbar\omega_{12}$. The vertical lines indicate four-wave mixing resonances as described in the main text. (b) Homodyne squeezing spectra as $1+\Lambda(\omega)$ with colours corresponding to the values of $\hbar\omega_{12}$ in panel (a). The vertical line indicates the sideband frequency $\hbar\omega=\frac{1}{2}(E^+_0-E^-_0)$, where two-mode squeezing is observed. (c) Zero-frequency squeezing $1+\Lambda(0)$ (solid line) and squeezing at the optimal homodyne frequency $1+\Lambda(\omega_{\rm opt})$ (dotted line) as a function of the drive frequency difference $\hbar\omega_{12}$. The vertical line indicates $\hbar\omega_{12} = \frac{1}{2}(E^+_0-E^-_0)$. (d) Fourier components $m=0$ and $m=2$ of parametric gain $\Delta^m$, shown as the sum of the norms of spin-diagonal and spin off-diagonal matrix elements. The full lines shows the total parametric gain, and the dashed lines show the contributions from the bound biexciton alone. The vertical line indicates $\hbar\omega_{12}=\frac{1}{2}(E^+_0-E^-_0)$.}
  \label{fig:config_B}
\end{figure*}

Fig.~\ref{fig:config_A}(c) shows the dependence of the squeezing at zero homodyne frequency as a function of the drive frequency difference $\hbar\omega_{12}$. As expected for Configuration A, the squeezing is strongest in the degenerate limit $\hbar\omega_{12}=0$, where both drives are resonant with the lower polariton. In addition, there is a weak resonance appearing at $\hbar\omega_{12}=E^+_0-E^-_0$, where the drive frequency $\omega_2$ is resonant with the upper polariton. The resonance is weak, because the energy of two upper polaritons overshoot the bound biexciton energy, $2E^+_0 \gg E^{\rm xx}_{\rm b,-}$.
We note that the bichromatic-pump scheme can be utilised to generate strong single-mode squeezing. Since the width of the squeezing dip around $\hbar\omega_{12}=0$ in Fig.~\ref{fig:config_A}(c) is several meV, one can choose a pump frequency difference of e.g. $\hbar\omega_{12} = 100\;{\rm \mu eV}$, which allows to filter out the pump lasers spectrally and still have strong single-mode squeezing over a bandwidth that exceeds the resolution of any standard photodetector.

In Fig.~\ref{fig:config_A}(d), the leading Fourier components $m=0$ and $m=2$ of the parametric gain $\bm{\Delta}^m$ is shown as the sum of the absolute values of the spin-diagonal and spin off-diagonal components. The $m=-2$ component  is simply the complex conjugate of $m=2$, and the remaining components are negligible in comparison to the ones shown. The reason behind this is twofold: first, the contributions to the parametric gain are products of $\ev*{a^\dagger}$ and $\ev{P^\dagger}$ or quantities that are driven by such products, and therefore only contain even Fourier orders. The Fourier components with $m$ larger than $2$ are only driven by higher-order processes and are therefore strongly suppressed in comparison with the leading components. This observation corroborates the validity of the DCT perturbative expansion up to third order in the pump field. However, we can not be entirely certain that the contributions to next (5th) order are of a different nature and can introduce new effects.
In addition to the total parameric gain, the contribution from the bound biexciton alone is also shown with dashed lines. As can be seen, this constitutes the dominating contribution to the parametric gain, and thus the other contributions in Eq.~\eqref{eq:Delta-matrix} from Pauli blocking and from the two-exciton scattering continuum are small corrections.

\subsection{Configuration B}
\label{sec:config-B}

Configuration B is defined by setting the cavity frequency such that $E^+_0+E^-_0=E^{\rm xx}_{\rm b,-}$ (corresponding to $E^{\rm p}_0=E^{\rm xx}_{\rm b,-}-E^{\rm x}_0$) and the mean driving frequency such that $\hbar\omega_1+\hbar\omega_2=E^+_0+E^-_0=E^{\rm xx}_{\rm b,-}$ [see Fig.~\ref{fig:1}(d)]. Thus, we expect the driving to be resonant, when the drive frequency difference $\hbar\omega_{12}=\frac{1}{2}(E^+_0-E^-_0)$, such that $\hbar\omega_{1}=E^-_0$ and $\hbar\omega_2=E^+_0$. In this case, the drive resonantly excites upper and lower polaritons, which scatter resonantly into bound biexcitons via the Coulomb interaction. The latter process is a \emph{non-degenerate} polaritonic Feshbach resonance, because the in-scattering particles (upper and lower polaritons) are different with non-degenerate energies.

Fig.~\ref{fig:config_B}(a) shows the spontaneous four-wave mixing spectra for Configuration B at different drive-frequency differences $\hbar\omega_{12}$. In contrast to Configuration A, the emission spectrum features two equally bright peaks from the upper and lower polariton, respectively. In terms of the four-wave mixing resonances introduced in Sec.~\ref{sec:config-A}, these peaks correspond to Eq.~\eqref{eq:w4-a} with $\hbar\omega_3=E^+_0$ and thus $\hbar\omega_4 = \hbar\omega_1+\hbar\omega_2-E^+_0=E^-_0$. These two main emission energies are indicated with grey vertical lines in Fig.~\ref{fig:config_B}(a). In addition, the weaker four-wave mixing resonance from Eq.~\eqref{eq:w4-c} $2\hbar\omega_2-E^-_0$ is indicated with vertical dotted lines. When the drive energy difference becomes larger, the power of the main emission peaks is weakened, whereby the additional four-wave mixing resonances $2\hbar\omega_2-E^+_0$ and $2\hbar\omega_1-E^+_0$ become visible as well (indicated with vertical dashed and dash-dotted lines, respectively).

In Fig.~\ref{fig:config_B}(b), the homodyne squeezing spectrum is shown. Here, the most remarkable difference from Configuration A is the appearance of maximal squeezing at the sideband frequency $\frac{1}{2}(E^+_0-E^-_0)$ rather than at zero homodyne frequency. This is because the non-degenerate Feshbach resonance creates two-mode squeezing, i.e. squeezing from the strong correlations between two different frequency bands. This is strongly related to the two equally strong peaks in the spontaneous four-wave mixing spectrum from Fig.~\ref{fig:config_B}(a) in contrast to Fig.~\ref{fig:config_A}(a), which features a single strong central peak. Although the detection of such high-frequency two-mode squeezing is challenging in standard homodyne detection due to finite detection bandwidth, succesful detection can be carried out with bichromatic heterodyne detection, where the high-frequency sideband squeezing is mixed down to a low-frequency beat signal~\cite{marino2007bichromatic,vaidya2020broadband}.

Fig.~\ref{fig:config_B}(c) shows the squeezing at zero homodyne frequency (solid line) and at the optimal homodyne sideband frequency (dotted line) as a function of the drive energy difference $\hbar\omega_{12}$. The optimal sideband frequency is close to $\frac{1}{2}(E^+_0-E^-_0)$, but due to small nonlinear shifts, we have taken the numerically optimal homodyne frequency. The behaviour shows a resonance around $\hbar\omega_{12}=\frac{1}{2}(E^+_0-E^-_0)$ (indicated with vertical line), where $\hbar\omega_1=E^-_0$ and $\hbar\omega_2=E^+_0$ and polaritons are efficiently excited by the drive fields. This resonance is also seen in Fig.~\ref{fig:config_B}(d), where the parametric gain is shown as a function of the drive energy difference $\hbar\omega_{12}$. Here, it is seen that the parametric gain is strongest around $\hbar\omega_{12}=\frac{1}{2}(E^+_0-E^-_0)$. Furthermore, as is the case for Configuration A, the bound biexciton (dashed lines) dominates the parametric gain in this configuration.

\section{Conclusion}
\label{sec:conclusion}

In conclusion, we have presented a theoretical investigation of spontaneous four-wave mixing and squeezing in bichromatically pumped atomically-thin semiconductor cavity polaritonic systems, in particular focusing on the strong nonlinear response from the bound biexciton. By applying a rigorous truncation scheme of the many-body state, we have derived a tractable set of equations of motions for exciton and photon fields, as well as the correlated multiparticle fields. In addition, we have employed a Heisenberg-Langevin approach to calculate the fluctuation spectra in the presence of two non-degenerate pump laser fields. The combination of these two methods gives access to the spontaneous four-wave mixing spectra and the squeezing properties of the outcoupled field from the cavity in the presence of strong Coulomb-generated correlations. We have focused on two resonant configurations, corresponding to a degenerate and non-degenerate polaritonic Feshbach resonance, respectively. In the degenerate configuration, a pair of lower polaritons is resonant with the bound biexciton, thereby giving rise to a single dominating peak in the spontaneous four-wave mixing spectrum and strong single-mode squeezing. In the non-degenerate configuration, a upper and lower polariton pair is resonant with the bound biexciton, thereby giving rise to two balanced peaks in the spontaneous four-wave mixing spectrum and strong two-mode squeezing. We believe that these results will open new opportunities in the cross-field between semiconductor physics and nonlinear optics.

Although the numerical calculations in this paper have been performed for an atomically-thin semiconductor, the overall features of the presented phenomena can be expected to be seen in other semiconductor materials with spectrally resolved biexcitons as well. We note that a previous experimental investigation~\cite{shimano2002efficient} measured parametric gain from biexcitons in a bulk CuCl microcavity in the near-UV spectral range, although no homodyne detection was performed in the experiment. Furthermore, ZnO quantum wells with biexciton binding energies around 15 meV~\cite{ko2000biexciton} are another interesting platform to potentially observe the predicted squeezing mechanism in the near-UV spectrum. Polariton Feshbach resonance has been observed in pump-probe experiments with InGaAs quantum wells~\cite{takemura2014polaritonic,takemura2017spin,navadeh2019polaritonic}.

\begin{acknowledgments}
E.V.D. acknowledges support from Independent Research Fund Denmark through an International Postdoc Fellowship (Grant No. 0164-00014B). A.K. and F.K. gratefully acknowledge support from the Deutsche Forschungsgemeinschaft through Projects No.~420760124 (\mbox{KN 427/11-1}) and No. 163436311---SFB~910 (Project B1).
\end{acknowledgments}

\appendix
\section{Matrix elements}
\label{sec:matrix-elements}
The equation of motion Eq.~\eqref{eq:eom} is derived as in Ref.~\onlinecite{denning2022efficient}, the only difference being the bichromatic input field as described in Sec.~\ref{sec:driving-field}. Here, we shall simply state the results from Ref.~\onlinecite{denning2022efficient}. The matrix elements in Eq.~\eqref{eq:eom} are given by
\begin{align}
\label{eq:exciton-eom-coefficients}
\begin{split}
  \tilde{\Omega}_\bq &= \sum_{\bk_1} A_\bq(
  \phi_{\bk_1}^*
  \phi_{\bk_1+\alpha\bq}
  \phi_{\bk_1 + \bq}
  +
  \phi_{\bk_1+\bq}^*
  \phi_{\bk_1+\alpha\bq}
  \phi_{\bk_1}
  \Big)
  \\
  W^0 &= \sum_{\bk_1\bk_2} V_{\bk_2-\bk_1}
  \phi_{\bk_1}\phi_{\bk_1}
  (\phi_{\bk_1}^*-\phi_{\bk_2}^*)
  (\phi_{\bk_1}^*-\phi_{\bk_2}^*)
  \\
  W^\pm_\mu &= \sum_\bq \Phi^\pm_{\mu,\bq} \tilde{W}^{\pm*}_{\bq,0}
  \\
  \widebar{W}^\pm_\mu &= \sum_{\bq\bq'} \widebar{\Phi}^\pm_{\mu,\bq}
  (\mathcal{S}^\pm)^{-1}_{\bq,\bq'}\tilde{W}^\pm_{\bq',0}
  \\
  \Omega^\pm_{\mu,\bq} &= \sum_{\bq'} \Phi^\pm_{\mu,-\bq'}\tilde{A}^\pm_{\bq',\bq}
  \\
  \widebar{\Omega}^\pm_{\mu,\bq} &= \sum_{\bq'\bq''}\widebar{\Phi}^\pm_{\mu,\bq'}
  (\mathcal{S}^{\pm})^{-1}_{\bq',\bq''} \tilde{A}^{\pm*}_{\bq'',\bq},
\end{split}
\end{align}
with
\begin{align}
\label{eq:two-momentum-coulomb-element}
  \begin{split}
  \tilde{W}_{\bq,\bq'}^\pm =
&\sum_{\bk_1\bk_2}
V_{\bq'-\bq}
\phi_{\bk_1}
\phi_{\bk_2}
[\phi^*_{\bk_1-\beta(\bq-\bq')} - \phi^*_{\nu_1,\bk_1+\alpha(\bq-\bq')}]
\\ &\hspace{3cm}\times
[\phi^*_{\bk_2+\beta(\bq-\bq')} - \phi^*_{\bk_2-\alpha(\bq-\bq')}]
\\
\pm &\sum_{\bk_1\bk_2}
V_{\bk_1-\bk_2+(\alpha-\beta)\bq+\bq'}
\phi_{\bk_1}
\phi_{\bk_2}
[\phi^*_{\bk_1-\beta(\bq-\bq')} - \phi^*_{\bk_2-\alpha(\bq+\bq')}]
\\
&\hspace{3cm}\times
[\phi^*_{\bk_1+\alpha(\bq+\bq')} - \phi^*_{\bk_2+\beta(\bq-\bq')}]
\\
\tilde{A}_{\bq',\bq}^\pm
&= \tilde{\Omega}_{\bq}\delta_{\bq\bq'}
  \mp A_\bq
  \sum_{\bk}\phi_{\bk + \alpha\bq}
  \phi_{\bk+\bq-\beta\bq'}^*
  \phi_{\bk-\alpha\bq'}^*
\\
\mathcal{S}_{\bq,\bq'}^\pm  &=
  \delta_{\bq\bq'}
  \mp
  \sum_{\bk}
  \phi_{\bk-\alpha\bq}
  \phi_{\bk+\bq'-\beta\bq}
  \phi_{\bk-\bq+\beta\bq'}^*
  \phi_{\bk+\alpha\bq'}^*.
\end{split}
\end{align}
The biexcitonic wavefunctions $\Phi^\pm_{\mu,\bq}$ are the solutions to the eigenvalue equation
\begin{align}
\label{eq:bix-eigenvalue}
  \qty(2E^{\rm x}_0 + \frac{\hbar^2 q^2}{M})\Phi^\pm_{\mu,\bq} + \sum_{\bq'\bq''}(\mathcal{S}^\pm)^{-1}_{\bq,\bq'}\tilde{W}^\pm_{\bq',\bq''} \Phi^\pm_{\mu,\bq''} = E^{\rm xx}_{\mu,\pm}\Phi^\pm_{\mu,\bq},
\end{align}
where $M = m_{\rm e}+m_{\rm h}$ is the total exciton mass. \rev{The numerical details of the solution of this eigenvalue equation is presented in Appendix~\ref{sec:appendix-numerical-calculations}}.

\section{Formal solutions of multiparticle fluctuations}
\label{sec:formal-solutions}

The first thing to notice is that the coupling coefficients between the correlated fluctuations $\delta\mathcal{B}$, $\delta\mathcal{C}$, and $\delta\mathcal{D}$ are linear and do not involve any steady-state expectation values. This means that the formal solution of these fluctuations follows the procedure in Ref.~\onlinecite{denning2022efficient} without any changes to the many-body quantities $\Pi$ and $K$. The only difference is the source terms involving $\delta P^\dagger$, $\delta a^\dagger$, $\delta a^{\rm in\dagger}$, $\delta P^{\rm in\dagger}$, which occur together with time-varying amplitudes. To incorporate this into the derivation, we first consider a general equation of motion involving two arbitrary fluctuation operators  $\delta Q$ and $\delta R$ as
\begin{align}
  -i\hbar\partial_t \delta Q(t) = A(t)\delta R(t)
\end{align}
where $A(t)$ is periodic with the period $T=\frac{2\pi}{\omega_{12}}$. We can then express $A(t)$ in terms of its discrete Fourier series as $A(t) = \sum_n A_n e^{-in\omega_{12}t}$ and write $\delta Q(t)=\frac{1}{2\pi}\int_{-\infty}^{\infty}\dd{\omega} e^{-i\omega t}\delta Q(\omega)$ and similarly for $\delta R$, such that
\begin{align}
\begin{split}
  -i\hbar\partial_t \frac{1}{2\pi}\int_{-\infty}^{\infty}\dd{\omega'} &e^{-i\omega' t}\delta Q(\omega')
  \\
  &= \sum_n A_n e^{-in\omega_{12}t}\int_{-\infty}^{\infty}\dd{\omega'} e^{-i\omega' t}\delta R(\omega').
\end{split}
\end{align}
Multiplying by $e^{i\omega t}$ and integrating over $t$, we find
\begin{align}
  -\hbar\omega\delta Q(\omega) = \sum_n A_n \delta R(\omega-n\omega_{12}).
\end{align}
Similarly, if we have a product of two amplitudes occurring on the right-hand side, we can use the properties of the Fourier series of a product,
\begin{align}
  C(t) = A(t)B(t) \Rightarrow C_m = \sum_n A_n B_{m-n}.
\end{align}
Thus a time-domain equation of motion of the form
$
  -i\hbar\partial_t \delta Q(t) = A(t)B(t)\delta R(t),
$
transforms to
$
  -\hbar\omega\delta Q(\omega) = \sum_m\sum_{n} A_n B_{m-n} \delta R(\omega-m\omega_{12}).
$.

With this, we can recycle the results from Ref.~\cite{denning2022efficient}, where the formal solution of the equation of motion for $\delta C$ [Eq. (S36) in the Supplementary Material of Ref.~\onlinecite{denning2022efficient}] becomes
%\begin{widetext}
\begin{align}
\label{eq:dC-formal-solution}
  \begin{split}
    \delta\mathcal{C}^{\zeta\zeta'}_{\bq}\!\!(\omega) &=
    i\hbar\sqrt{2\gamma^{\rm p}}\sum_{\zeta_1\zeta_1'}
    K^{\zeta\zeta'\bq}_{\zeta_1\zeta_1'0}(\omega)
    \sum_n\ev*{a^{\rm in\dagger}_{\zeta_1}}_n\delta P^\dagger_{\zeta_1',0}(\omega-n\omega_{12})
    \\
    &\hspace{-1.4cm}-i\hbar\sqrt{2\gamma^{\rm p}}\sum_{\zeta_1\zeta_1'}
    K^{\zeta\zeta'\bq}_{\zeta_1\zeta_1'0}(\omega)
    \frac{\Omega_0}{\hbar\omega+2(\tilde{E}^{\rm p}_0-\hbar\omega_{\rm d})}
    \\
    &\hspace{-1.4cm}\times\sum_n\Big[
    \ev*{a^{\rm in\dagger}_{\zeta_1}}_n\delta a^\dagger_{\zeta_1',0}(\omega-n\omega_{12})
    +\ev*{a^{\rm in\dagger}_{\zeta_1'}}_n\delta a^\dagger_{\zeta_1,0}(\omega-n\omega_{12})
    %\Big]
    %\\
    %&-i\hbar\sqrt{2\gamma^{\rm p}}\sum_{\zeta_1\zeta_1'}
    %K^{\zeta\zeta'\bq}_{\zeta_1\zeta_1'0}(\omega)
    %\frac{\Omega_0}{\hbar\omega+2(\tilde{E}^{\rm p}_0-\hbar\omega_{\rm d})}
    %\sum_n\qty[
    \\ &\hspace{-0.5cm}
    +\ev*{a^\dagger_{\zeta_1,0}}_n\delta a^{\rm in\dagger}_{\zeta_1'}(\omega-n\omega_{12})
    + \ev*{a^\dagger_{\zeta_1',0}}_n\delta a^{\rm in\dagger}_{\zeta_1}(\omega-n\omega_{12})
    \Big]
    \\ &\hspace{-1.4cm}
    + i\hbar\sqrt{2\gamma^{\rm x}}
    \sum_{\zeta_1\zeta_1'\bq_1}
    K^{\zeta\zeta'\bq}_{\zeta_1\zeta_1'\bq_1}\!\!(\omega)
    \sum_n
    \Bigg\{\delta_{\bq_1,0}\ev*{a^\dagger_{\zeta_1,0}}_n\delta P^{\rm in\dagger}_{\zeta_1',0}(\omega-n\omega_{12})
    \\
    &\hspace{-1.4cm}-\sum_{\mu\pm}
    \frac{\frac{1}{4}(1\pm\delta_{\zeta_1\zeta_1'})\widebar{\Phi}^\pm_{\mu,0}\Omega^\pm_{\mu,\bq_1}}{\hbar\omega+\tilde{E}^{\rm xx}_{\mu,\pm}-2\hbar\omega_{\rm d}}
    \Big[\ev*{P^\dagger_{\zeta_1,0}}_n\delta P^{\rm in\dagger}_{\zeta_1',0}(\omega-n\omega_{12})
    \\
    &\hspace{2.5cm}+ \ev*{P^\dagger_{\zeta_1',0}}_n\delta P^{\rm in\dagger}_{\zeta_1,0}(\omega-n\omega_{12})
    \Big]\Bigg\},
  \end{split}
\end{align}
%\end{widetext}
where
\begin{align}
\begin{split}
K(\omega) = -\Big[
&\delta_{\bq,\bq_1}\delta_{\zeta\zeta_1}\delta_{\zeta',\zeta_1'}
(\hbar\omega + \tilde{E}^{\rm x}_\bq + \tilde{E}^{\rm p}_\bq-2\hbar\omega_{\rm d})
\\
&\hspace{3cm}+ \Pi^{\zeta\zeta'\bq}_{\zeta_1\zeta_1'\bq_1}(\omega)
\Big]^{-1}.
\end{split}
\end{align}
is the Green's function for $\delta\mathcal{C}(\omega)$ with self-energy
\begin{align}
\begin{split}
  \Pi^{\zeta\zeta'\bq}_{\zeta_1\zeta_1'\bq_1}\!\!(\omega) &=
  -\frac{\Omega_\bq\Omega_{-\bq_1}}{\hbar\omega+2(\tilde{E}^{\rm p}_\bq-\hbar\omega_{\rm d})}
  \\
  &\times\qty[\delta_{\zeta',\zeta_1}\delta_{\zeta,\zeta_1'}\delta_{-\bq,\bq_1}
  + \delta_{\zeta,\zeta_1}\delta_{\zeta',\zeta_1'}\delta_{\bq,\bq_1}]
  \\ &\hspace{-1cm}
  -\sum_{\mu\pm}
  \frac{\frac{1}{2}(1\pm\delta_{\zeta\zeta'})\Omega^\pm_{\mu,\bq} \widebar{\Omega}_{\mu,\bq_1}^\pm}{\hbar\omega+\tilde{E}^{\rm xx}_{\mu,\pm}-2\hbar\omega_{\rm d}}
  [\delta_{\zeta'\zeta_1}\delta_{\zeta,\zeta_1'}
  + \delta_{\zeta\zeta_1}\delta_{\zeta'\zeta_1'}].
\end{split}
\end{align}
The Fourier-transformation of the equation of motion for $\delta P^\dagger$, Eq.~\eqref{eq:fluctuation-eom-1}, becomes
\begin{align}
\label{eq:exciton-fluctuation-eom-2}
  \begin{split}
    -\hbar\omega&\delta P_{\zeta,0}^\dagger(\omega)
    = \tilde{E}_0^{\rm x} \delta P_{\zeta,0}^\dagger(\omega)
    + \Omega_0\delta a_{\zeta,0}^\dagger(\omega)
    \\
    &+\sum_{\zeta'}\sum_n\Delta_{\zeta\zeta'}^n\delta P_{\zeta',0}(\omega-n\omega_{12})
    + i\hbar\sqrt{2\gamma^{\rm x}} \delta P_{\zeta,0}^{\rm in\dagger}(\omega),
    \\
    &+ \sum_{\zeta'\zeta_1\zeta_2\bq}\sum_n \ev*{P_{\zeta',0}}_n Q^{\zeta\zeta'}_{\zeta_1\zeta_2\bq}(\omega-n\omega_{12})
    \delta\mathcal{C}^{\zeta_1\zeta_2}_\bq(\omega-n\omega_{12})
    \\
    &- i\hbar\sqrt{2\gamma^{\rm x}}
    \frac{1}{2}\sum_{\zeta'\mu\pm}\sum_{nn'}
    \ev*{P_{\zeta',0}}_{n}
    \frac{\frac{1}{2}(1\pm\delta_{\zeta\zeta'})W^\pm_\mu\widebar{\Phi}^\pm_{\mu,0}}
    {\hbar(\omega-n\omega_{12})+\tilde{E}^{\rm xx}_{\mu,\pm}-2\hbar\omega_{\rm d}}
    \\ &\hspace{2cm}\times
    \Big[\ev*{P^\dagger_{\zeta,0}}_{n'}\delta P^{\rm in\dagger}_{\zeta',0}(\omega-[n+n']\omega_{12})
    \\ &\hspace{2.5cm}+ \ev*{P^\dagger_{\zeta',0}}_{n'}\delta P^{\rm in\dagger}_{\zeta,0}(\omega-[n+n']\omega_{12})
    \Big],
  \end{split}
\end{align}
where $\Delta^m_{\zeta\zeta'}$ is defined in Eq.~\eqref{eq:Delta-matrix},
and where the $Q$-matrix is defined slightly different than in Ref.~\onlinecite{denning2022efficient},
\begin{align}
\begin{split}
Q^{\zeta\zeta'}_{\zeta_1\zeta_2\bq}\!(\omega) &= -
\delta_{\zeta\zeta'}\delta_{\zeta_1\zeta}\delta_{\zeta_2\zeta}\tilde{\Omega}_\bq
\\ &-
\sum_{\mu\pm}
\frac{\frac{1}{2}(1\pm\delta_{\zeta_1\zeta_2})W^\pm_\mu\widebar{\Omega}_{\mu,\bq}^\pm
(\delta_{\zeta_2\zeta}\delta_{\zeta_1\zeta'} + \delta_{\zeta_1\zeta}\delta_{\zeta_2\zeta'})}
{\hbar\omega+\tilde{E}^{\rm xx}_{\mu,\pm}-2\hbar\omega_{\rm d}}.
\end{split}
\end{align}

Inserting the formal solution for $\delta \mathcal{C}$ into Eq.~\eqref{eq:exciton-fluctuation-eom-2} and writing all frequency-dependent quantities using the Fourier-index form, we find
\begin{align}
\begin{split}
  -\sum_{\zeta'm'}&\qty{[\hbar(\nu-m\omega_{12}) + \tilde{E}^{\rm x}_0]\delta_{\zeta\zeta'}\delta_{mm'} + \Sigma_{\zeta\zeta'}^{mm'}(\nu)}\delta P^\dagger_{\zeta'm'}(\nu)
  \\
  &= \sum_{\zeta'm'} \Omega_{\zeta\zeta',0}^{mm'}(\nu) \delta a_{\zeta',m'}^\dagger(\nu)
  +\sum_{\zeta'm'}\Delta_{\zeta\zeta'}^{m'-m}\delta P_{\zeta',m'}(\nu)
  \\ &+ \sum_{\zeta'm'} T^{{\rm x},mm'}_{\zeta\zeta'}(\nu) \delta P^{\rm in\dagger}_{\zeta'm'}(\nu)
  + \sum_{\zeta'm'} T^{{\rm p},mm'}_{\zeta\zeta'}(\nu) \delta a^{\rm in\dagger}_{\zeta'm'}(\nu),
\end{split}
\end{align}
where the self-energy and renormalized coupling matrices are given by

\begin{align}
\begin{split}
  \Sigma_{\zeta\zeta'}^{mm'}(\nu) &= i\hbar\sqrt{2\gamma^{\rm p}}
  \sum_{nn'}\sum_{\zeta_1\zeta_2\bq}\sum_{\zeta_1'\zeta_2'}
  \ev*{P_{\zeta_2',0}}_n Q^{\zeta\zeta_2'}_{\zeta_1\zeta_2\bq,m+n}(\nu)
  \\ &\times
  K^{\zeta_1\zeta_2\bq}_{\zeta_1'\zeta'0,m+n}(\nu)\ev*{a^{\rm in\dagger}_{\zeta_1'}}_{n'}
  \delta_{m',m+n+n'}
  \\
  \hat{\Omega}^{mm'}_{\zeta\zeta'0}(\nu) &=
  \Omega_0 \delta_{mm'}\delta_{\zeta\zeta'}
  - i\hbar\sqrt{2\gamma^{\rm p}}\sum_{nn'}\sum_{\zeta_1\zeta_2\zeta_3\bq}\sum_{\zeta_1'\zeta_2'}
  \\ &\times
  \ev*{P_{\zeta_3,0}}_n
  \frac{\Omega_0 Q^{\zeta\zeta_3}_{\zeta_1\zeta_2\bq,m+n}(\nu)
  K^{\zeta_1\zeta_2\bq}_{\zeta_1'\zeta_2'0,m+n}(\nu)}{\hbar(\nu-[m+n]\omega_{12}) + 2(\tilde{E}^{\rm p}_0-\hbar\omega_{\rm d})}
  \\ &\times
  [\ev*{a^{\rm in\dagger}_{\zeta_1'}}_{n'}\delta_{\zeta'\zeta_2'}
  + \ev*{a^{\rm in\dagger}_{\zeta_2'}}_{n'}\delta_{\zeta'\zeta_1'}]
  \delta_{m',m+n+n'}
  \end{split}
  \end{align}
and the renormalized incoupling matrices are given by
  \begin{align}
  \begin{split}
  T^{{\rm p},mm'}_{\zeta\zeta'}(\nu) &=
  i\hbar\sqrt{2\gamma^{\rm p}}\sum_{nn'}\sum_{\zeta_1\zeta_2\zeta_3\bq}\sum_{\zeta_1'\zeta_2'}
  \ev*{P_{\zeta_3,0}}_n Q^{\zeta\zeta_3}_{\zeta_1\zeta_2\bq,m+n}(\nu)
  \\
  &\times
  K^{\zeta_1\zeta_2\bq}_{\zeta_1'\zeta_2'0,m+n}(\nu)
  \Bigg\{\ev*{P_{\zeta_2',0}^\dagger}_{n'}\delta_{\zeta'\zeta_1'}
  \\
  &-
  \frac{\Omega_0[\ev*{a^{\dagger}_{\zeta_1',0}}_{n'}\delta_{\zeta'\zeta_2'}
  + \ev*{a^{\dagger}_{\zeta_2',0}}_{n'}\delta_{\zeta'\zeta_1'}]}{\hbar(\nu-[m+n]\omega_{12}) + 2(\tilde{E}^{\rm p}_0-\hbar\omega_{\rm d})}
  \Bigg\}
  \delta_{m',m+n+n'}
  \\
  T^{{\rm x},mm'}_{\zeta\zeta'}(\nu) &= i\hbar\sqrt{2\gamma^{\rm x}}\Bigg\{
  \delta_{\zeta\zeta'}\delta_{mm'}
  -
  \frac{1}{2}\sum_{\zeta_1\mu\pm}\sum_{nn'}
  \ev*{P_{\zeta',0}}_{n}
  \\ &\times
  \frac{\frac{1}{2}(1\pm\delta_{\zeta\zeta_1})W^\pm_\mu\widebar{\Phi}^\pm_{\mu,0}}
  {\hbar(\nu-[m+n]\omega_{12})+\tilde{E}^{\rm xx}_{\mu,\pm}-2\hbar\omega_{\rm d}}
  \\ &\times
  \qty[\ev*{P^\dagger_{\zeta,0}}_{n'}\delta_{\zeta'\zeta_1}
  + \ev*{P^\dagger_{\zeta_1,0}}_{n'}\delta_{\zeta'\zeta}]\delta_{m',m+n+n'}
  \Bigg\}
  \\ &+i\hbar\sqrt{2\gamma^{\rm x}}
  \sum_{nn'}\sum_{\zeta_1\zeta_2\zeta_3\bq}\sum_{\zeta_1'\zeta_2'}
  \ev*{P_{\zeta_3,0}}_n Q^{\zeta\zeta_3}_{\zeta_1\zeta_2\bq,m+n}(\nu)
  \\ &\times
  K^{\zeta_1\zeta_2\bq}_{\zeta_1'\zeta_2'\bq_1,m+n}(\nu)
  \Bigg\{\delta_{\bq_1,0}\ev*{a^\dagger_{\zeta_1',0}}_{n'}\delta_{\zeta'\zeta_2'}
  \\
  &-\sum_{\mu\pm}
  \frac{\frac{1}{4}(1\pm\delta_{\zeta_1'\zeta_2'})\widebar{\Phi}^\pm_{\mu,0}\Omega^\pm_{\mu,\bq_1}}{\hbar(\nu-[m+n]\omega_{12})+\tilde{E}^{\rm xx}_{\mu,\pm}-2\hbar\omega_{\rm d}}
  \\ &\times
  \qty[\ev*{P^\dagger_{\zeta_1',0}}_{n'}\delta_{\zeta'\zeta_2'}
  + \ev*{P^\dagger_{\zeta_2',0}}_{n'} \delta_{\zeta'\zeta_1'}]\Bigg\}\delta_{m',m+n+n'},
\end{split}
\end{align}
with $K^{\zeta_1\zeta_2\bq}_{\zeta_1'\zeta'0,m}(\nu) := K^{\zeta_1\zeta_2\bq}_{\zeta_1'\zeta'0}(\nu-m\omega_{12})$ and similarly for $Q^{\zeta\zeta_2'}_{\zeta_1\zeta_2\bq,m}(\nu)$.

\section{Homodyne noise spectrum}
\label{sec:appendix-homodyne}
In this appendix, we provide additional details of the derivation of the homodyne noise spectrum.

In terms of the fields $E_{\rm lo}$ and $E_{\rm s}$ given in Eq.~\eqref{eq:homodyne-beamsplitter}, the mean photocurrent difference from Eq.~\eqref{eq:mean-photocurrent} takes the form
\begin{align}
\label{eq:mean-photocurrent-2}
\begin{split}
\overline{I_-(t)}&:=\overline{I_1(t)}-\overline{I_2(t)} \\ &=i\frac{\eta}{\tau_{\rm d}}\int_{t-\tau_{\rm d}}^t\dd{t'} \Big[\ev*{E_{\rm lo}(t')}\ev*{E_{\rm s}^\dagger(t')}
- \ev*{E_{\rm lo}^\dagger(t')}\ev*{E_{\rm s}(t')}\Big]
\\
&=\frac{\eta\sqrt{F_{\rm lo}}}{\tau_{\rm d}}\int_{t-\tau_{\rm d}}^t\dd{t'} \ev*{X_{\rm s}(\theta,t')},
\end{split}
\end{align}
where $X_{\rm s}(\theta,t) = e^{i\theta} E_{\rm s}^\dagger(t) + e^{-i\theta}E_{\rm s}(t)$ is the source-field quadrature operator with angle $\theta=\varphi+\pi/2$. The two terms in $\theta$ stem from the local oscillator phase $\varphi$ and a phase displacement $\pi/2$ from the beamsplitter.

Similarly, the correlation function of the photocurrent difference from Eq.~\eqref{eq:photocurrent-correlation-1} becomes
\begin{align}
  \label{eq:photocurrent-correlation-2}
\begin{split}
  &\overline{I_-(t)I_-(t+\tau)} =
  \overline{I_1(t)I_1(t+\tau)} + \overline{I_2(t)I_2(t+\tau)} \\
  &\hspace{4cm}- \overline{I_1(t)I_2(t+\tau)} - \overline{I_2(t)I_1(t+\tau)} \\
&
=\frac{1}{\tau_{\rm d}^2}\Bigg\{\eta^2\int_{t-\tau_{\rm d}}^t\dd{t'}\int_{t+\tau-\tau_{\rm d}}^{t+\tau}\dd{t''} \Big[
\ev*{E_{\rm lo}^{\dagger}(t')E_{\rm lo}(t'')}\ev*{E_{\rm s}^\dagger(t'')E_{\rm s}(t')}
\\
&\hspace{3cm}
+\ev*{E_{\rm lo}^\dagger(t'') E_{\rm lo}(t')}\ev*{E_{\rm s}^\dagger(t')E_{\rm s}(t'')}
\\
&\hspace{3cm}
-\ev*{:E_{\rm lo}^\dagger(t') E_{\rm lo}^\dagger(t''):}\ev*{:E_{\rm s}(t') E_{\rm s}(t''):}
\\
&\hspace{3cm}
-\ev*{:E_{\rm lo}(t') E_{\rm lo}(t''):}\ev*{:E_{\rm s}^\dagger(t') E_{\rm s}^\dagger(t''):}
\Big] \\
&
+\eta\Theta(\tau_{\rm d}-\tau)\int_{t+\tau-\tau_{\rm d}}^t\dd{t'} \ev*{E_{\rm s}^\dagger(t')E_{\rm s}(t')} + \ev*{E_{\rm lo}^\dagger(t')E_{\rm lo}(t')}.
\Bigg\}
\end{split}
\end{align}
The last term is the shot noise correlation function, $N_0(t,\tau)=(\eta/\tau_{\rm d}^2)\Theta(\tau_{\rm d}-\tau)\int_{t+\tau-\tau_{\rm d}}^t\dd{t'} [\ev*{E_{\rm s}^\dagger(t')E_{\rm s}(t')} + \ev*{E_{\rm lo}^\dagger(t')E_{\rm lo}(t')}]$. In typical homodyne detection setups, the local oscillator is significantly stronger than the signal, which means that the shot noise will be dominated by the local oscillator. In this limit, we can neglect the signal contribution to $N_0(t,\tau)$ which then becomes
\begin{align}
  N_0(t,\tau) \simeq N_0(\tau) = \frac{\eta F_{\rm lo}}{\tau_{\rm d}^2}(\tau_{\rm d}-\tau)\Theta(\tau_{\rm d}-\tau).
\end{align}
Combining Eqs.~\eqref{eq:mean-photocurrent-2} and \eqref{eq:photocurrent-correlation-2}, we find
\begin{align}
\label{eq:photocurrent-correlation-4}
\begin{split}
  &\overline{I_-(t)I_-(t+\tau)} -
  \overline{I_-(t)}\;\;\overline{I_-(t+\tau)}
\\
  &= N_0(\tau) + \frac{\eta^2F_{\rm lo}}{\tau_{\rm d}^2}
  \int_{t-\tau_{\rm d}}^t\dd{t'}\int_{t+\tau-\tau_{\rm d}}^{t+\tau}\dd{t''} \ev*{:\delta X_{\rm s}(\theta,t')\delta X_{\rm s}(\theta,t''):},
\end{split}
\end{align}
where $\delta X_{\rm s}(\theta,t) := X_{\rm s}(\theta,t)-\ev*{X_{\rm s}(\theta,t)}$. In the second integral, we make the substitution $t'''= t''-\tau$ in order to make the integration limits equal. Thereby, we obtain after relabelling $t'''$ to $t''$
\begin{align}
\begin{split}
  &\overline{I_-(t)I_-(t+\tau)} -
  \overline{I_-(t)}\;\;\overline{I_-(t+\tau)}
\\
  &= N_0(\tau) + \frac{\eta^2F_{\rm lo}}{\tau_{\rm d}^2}
  \int_{t-\tau_{\rm d}}^t\!\!\!\!\!\!\dd{t'}
  \int_{t-\tau_{\rm d}}^{t}\!\!\!\!\!\!\dd{t''}
  \ev*{:\delta X_{\rm s}(\theta,t')\delta X_{\rm s}(\theta,t''+\tau):}.
\end{split}
\end{align}

Inserting this expression into Eq.~\eqref{eq:Nw-definition}, we obtain the noise spectrum
\begin{align}
\label{eq:Nw-1}
\begin{split}
  N(\omega)
 &= N_0(\omega) + \frac{\eta^2F_{\rm lo}}{\tau_{\rm d}^2}
\lim_{t_0\rightarrow\infty}\frac{1}{T}\int_{t_0}^{t_0+T}\dd{t}\int_0^\infty\dd{\tau} \cos(\omega\tau)
\\ &\times
\int_{t-\tau_{\rm d}}^t\!\!\!\!\!\!\dd{t'}
\int_{t-\tau_{\rm d}}^{t}\!\!\!\!\!\!\dd{t''}
\\ &\times\Big\{
\ev*{\delta E_{\rm s}^\dagger (t')\delta E_{\rm s}(t''+\tau)}
+ \ev*{\delta E_{\rm s}^\dagger(t''+\tau) \delta E_{\rm s}(t')}
\\
&\hspace{1cm}+ e^{-2i\theta}[
\Theta(t'-t''-\tau)\ev*{\delta E_{\rm s}(t')\delta E_{\rm s}(t''+\tau)}
\\
&\hspace{2cm}+ \Theta(t''+\tau-t')\ev*{\delta E_{\rm s}(t''+\tau)\delta E_{\rm s}(t')}
]
\\ &\hspace{1cm}
+ e^{2i\theta}[
\Theta(t''+\tau-t')\ev*{\delta E_{\rm s}^\dagger(t')\delta E_{\rm s}^\dagger(t''+\tau)}
\\
&\hspace{2cm}+ \Theta(t'-t''-\tau)\ev*{\delta E_{\rm s}^\dagger(t''+\tau)\delta E_{\rm s}^\dagger(t')}
]\Big\}
\end{split}
\end{align}
where $N_0(\omega) = \lim_{t_0\rightarrow\infty}\frac{1}{T}\int_{t_0}^{t_0+T}\dd{t}\int_0^\infty\dd{\tau} \cos(\omega\tau) N_0(\tau)
= \frac{1}{2}\eta F_{\rm lo} \frac{\sin^2(\omega\tau_{\rm d}/2)}{(\omega\tau_{\rm d}/2)^2}$ is the contribution from shot noise to the homodyne noise spectrum. The quantity $\sin^2(\omega\tau_{\rm d}/2)/(\omega\tau_{\rm d}/2)^2$ is a filter factor that arises from the $\tau$-integral and describes the bandwidth of the detector as the inverse response time $\tau_{\rm d}^{-1}$.

The first term in Eq.~\eqref{eq:Nw-1} can be rewritten using the spectral correlation function $S_1^{mm'}(\nu)$ from Eq.~\eqref{eq:S12-spectral-correlation-functions} as
\begin{align}
\label{eq:Lambda1-1}
  \begin{split}
  &\mathcal{N}_1 := \lim_{t_0\rightarrow\infty}\frac{1}{T}\int_{t_0}^{t_0+T}\dd{t}
  \int_{t-\tau_{\rm d}}^t\!\!\!\!\!\!\dd{t'}
  \int_{t-\tau_{\rm d}}^{t}\!\!\!\!\!\!\dd{t''}
  \ev*{\delta E_{\rm s}^\dagger (t')\delta E_{\rm s}(t''+\tau)}
  \\
  &= \frac{1}{T}\int_{t_0}^{t_0+T}\dd{t}
  \int_{t-\tau_{\rm d}}^t\!\!\!\!\!\!\dd{t'}
  \int_{t-\tau_{\rm d}}^{t}\!\!\!\!\!\!\dd{t''}
  \frac{1}{(2\pi)^2}
  \int_{-\infty}^\infty\dd{\omega'}
  \int_{-\infty}^\infty\dd{\omega''}
  \\ &\times
  e^{i\omega't'}e^{-i\omega''(t''+\tau)}
  \ev*{\delta E_{\rm s}^\dagger (-\omega')\delta E_{\rm s}(\omega'')}
  \\ &=
  \frac{1}{T}\int_{t_0}^{t_0+T}\dd{t}
  \int_{t-\tau_{\rm d}}^t\!\!\!\!\!\!\dd{t'}
  \int_{t-\tau_{\rm d}}^{t}\!\!\!\!\!\!\dd{t''}
  \frac{1}{2\pi}
  \int_{-\omega_{12}/2}^{\omega_{12}/2}\!\!\!\!\!\!\dd{\nu'}
  \int_{-\omega_{12}/2}^{\omega_{12}/2}\!\!\!\!\!\!\dd{\nu''}
  \sum_{m'm''}
  \\ &\times
  e^{i(\nu'-m'\omega_{12})t'}e^{-i(\nu''-m''\omega_{12})(t''+\tau)}
  \delta(\nu'-\nu'')2\gamma^{\rm p}S^{m',m''}_1(\nu'),
  \end{split}
\end{align}
where we used the identity $\int_{-\infty}^\infty\dd{\omega}=\sum_m\int_{-\omega_{12}/2}^{\omega_{12}/2}\dd{\nu}$.
By shifting the $t'$ and $t''$ integration variables to $t'-t$ and $t''-t$, the integration over $t$ can be carried out, which gives the kronecker delta $\frac{1}{T}\int_{t_0}^{t_0+T}\dd{t} e^{i(m''-m')\omega_{12}t} = \delta_{m'm''}$.
Upon resolving this Kronecker delta as well as the delta function $\delta(\nu'-\nu'')$ and subsequently rewriting the summation over $\nu'$ and $m'$ to an integral over $\omega'$, we can resolve the integrals over $t'$ and $t''$, leading to
\begin{align}
  \begin{split}
    \mathcal{N}_1 &=
    \frac{1}{2\pi}\int_{-\infty}^\infty\dd{\omega'} \frac{\sin^2(\omega'\tau_{\rm d}/2)}{(\omega'/2)^2}e^{-i\omega'\tau} 2\gamma^{\rm p}S_1(\omega').
  \end{split}
\end{align}
The second term in Eq.~\eqref{eq:Nw-1}, where the time arguments are reversed, yields the complex conjugate $\mathcal{N}_1^*$.

For the term in Eq.~\eqref{eq:Nw-1} proportional to $e^{-2i\theta}$, the situation is slightly more complicated because of the Heaviside functions. Similarly to Eq.~\eqref{eq:Lambda1-1}, we can express the temporal correlation function in terms of $S_2^{mm'}(\nu)$ and resolve the integral over $t$, such that
\begin{align}
\label{eq:Lambda2-1}
  \begin{split}
  &\mathcal{N}_2 = \lim_{t_0\rightarrow\infty}\frac{1}{T}
  \int_{t_0}^{t_0+T}\!\!\!\!\!\!\dd{t}
  \int_{t-\tau_{\rm d}}^t\!\!\!\!\!\!\dd{t'}
  \int_{t-\tau_{\rm d}}^{t}\!\!\!\!\!\!\dd{t''}
  \\ &\hspace{2cm}\times
  [\Theta(t'-t''-\tau)\ev*{\delta E_{\rm s}(t')\delta E_{\rm s}(t''+\tau)}
  \\
  &\hspace{3cm}+ \Theta(t''+\tau-t')\ev*{\delta E_{\rm s}(t''+\tau)\delta E_{\rm s}(t')}]
  \\
  &= \int_{-\tau_{\rm d}}^0\!\!\!\!\!\!\dd{t'}
  \int_{-\tau_{\rm d}}^0\!\!\!\!\!\!\dd{t''}
  \frac{1}{2\pi}\int_{-\infty}^\infty\dd{\omega'}
  [\Theta(t'-t''-\tau)e^{i\omega't'}e^{-i\omega''(t''+\tau)}
  \\ &\hspace{2cm}+
  \Theta(t''+\tau-t')e^{-i\omega''t'}e^{i\omega'(t''+\tau)}]
  2\gamma^{\rm p}S_2(\omega').
  \end{split}
\end{align}
To resolve the integrals over $t'$ and $t''$, we now make the assumption that the detector response time $\tau_{\rm d}$ is much faster than any time scale in the dynamics of the source field. This corresponds to taking the limit of infinite detection bandwidth. Naturally, the detection bandwidth will restrict the observable homodyne spectrum, but in the interest of characterizing the intrinsic properties of the source field, $\tau_{\rm d}\rightarrow 0$ is the correct limit to consider. Since the integral over $\tau$ in Eq.~\eqref{eq:Nw-1} only runs over positive values, this means that only the second Heaviside function in Eq.~\eqref{eq:Lambda2-1} can be nonzero. Thus, we end up with (for $\tau>0$)
\begin{align}
  \mathcal{N}_2 &= \frac{1}{2\pi}
  \int_{-\infty}^\infty\dd{\omega'} \frac{\sin^2(\omega'\tau_{\rm d}/2)}{(\omega'/2)^2}e^{i\omega'\tau}2\gamma^{\rm p}S_2(\omega').
\end{align}
The term in Eq.~\eqref{eq:Nw-1} proportional to $e^{2i\theta}$ can be connected to $S_2^*(\omega)$ by following an analogous derivation. The final expression for the quadrature noise spectrum becomes
\begin{align}
  N(\omega) = \frac{1}{2}\eta F_{\rm lo}
  \frac{\sin^2(\omega\tau_{\rm d}/2)}{(\omega\tau_{\rm d}/2)^2}
  \{1 + \eta\Re[\Lambda_1(\omega) + e^{-2i\theta}\Lambda_2(\omega)]\},
\end{align}
where the functions $\Lambda_i(\omega)$ are defined below Eq.~\eqref{eq:Nw-2} in the main text.
In order to characterize the intrinsic noise properties of the source field, we set the detection efficiency $\eta$ to unity. Thereby, when normalizing the quadrature noise spectrum to the shot-noise level, we obtain
\begin{align}
  \frac{N(\omega)}{N_0(\omega)} = 1 +\Re[\Lambda_1(\omega) + e^{-2i\theta}\Lambda_2(\omega)].
\end{align}

\section{Effect of multiparticle fluctuation renormalization}
\label{sec:appendix-multiparticle-effect}
\begin{figure}[h]
\includegraphics[width=\columnwidth]{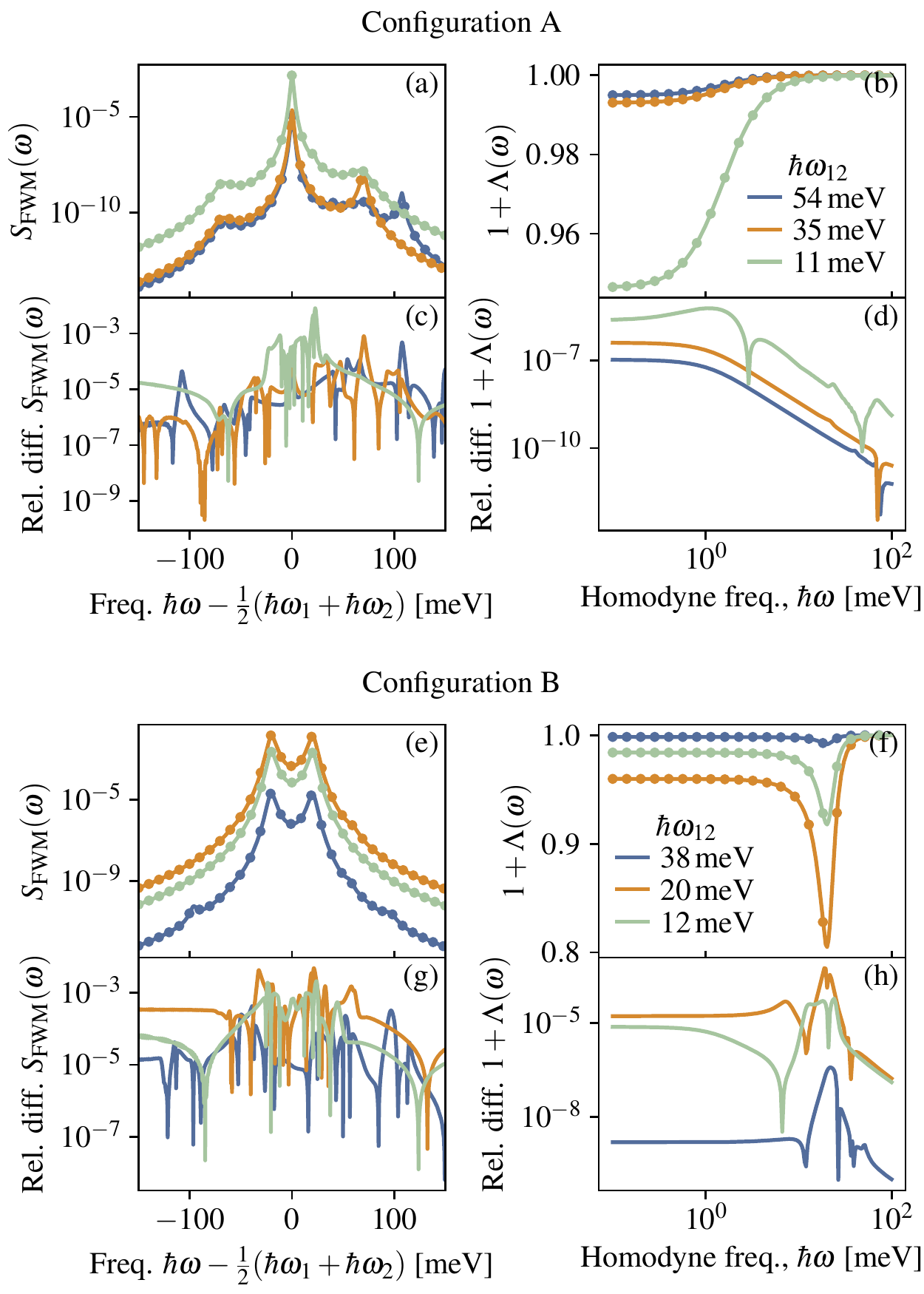}
  \caption{(a)--(b) Comparison of spontaneous four-wave mixing spectrum $S_{\rm FWM}$ and squeezing spectrum $1+\Lambda(\omega)$ from the full theoretical calculations (solid lines) to the simplifications from Eq.~\eqref{eq:DCTA-simplifications} (dots) for Configuration A and with the same parameter settings as for Fig.~\ref{fig:config_A}. (c)--(d) Absolute relative deviation between the full and simplified calculations. (e)--(f) Same as (a)--(b) for Configuration B, and with parameter settings as in Fig.~\ref{fig:config_B}. (g)--(h) Same as (c)--(d) for Configuration B. }
  \label{fig:DCTA}
\end{figure}

As mentioned in the main text, the multiparticle renormalizations of the Heisenberg-Langevin equations of motion only give small quantitative corrections to the spontaneous four-wave mixing and squeezing spectra. This is explicitly demonstrated in Fig.~\ref{fig:DCTA}, which compares the full and simplified calculations for the same parameters as in Figs.~\ref{fig:config_A} and \ref{fig:config_B}. Specifically, Fig.~\ref{fig:DCTA}(a)--(b) show the spontaneous four-wave mixing spectrum and squeezing spectrum calculated from the full theory (solid lines) and using the simplifications in Eq.~\eqref{eq:DCTA-simplifications} (dots). In Fig.~\ref{fig:DCTA}(c)--(d), the relative deviation between the two calculations are shown. This relative deviation $\delta_x$ is calculated as $\delta_x(\omega)=\abs{x_{\rm full}(\omega) - x_{\rm simplified}(\omega)}/x_{\rm full}(\omega)$, where $x(\omega)$ is either $S_{\rm FWM}(\omega)$ or $1+\Lambda(\omega)$. Evidently, the relative deviation is below 1\% for the spontaneous four-wave mixing spectrum, and below $10^{-5}$ for the squeezing spectrum.

\section{Numerical calculations}
\label{sec:appendix-numerical-calculations}
The numerical calculations in the paper have been performed for atomically thin $\mathrm{MoS_2}$ encapsulated with hexagonal BN on both sides, which is often used in experiments [see e.g.~\cite{sidler2017fermi}]

We use the screened Coulomb potential obtained from solving Poison's equation for the van der Waals heterostructure: dielectric environment/air gap/atomically thin semiconductor/air gap/dielectric environment \cite{florian2018dielectric,steinhoff2020dynamical}.
The small interlayer air gaps $h_{\rm int}$ take account of naturally occurring air gaps between the atomically thin semiconductor and its dielectric environment \cite{rooney2017observing} described by the dielectric constant $\varepsilon_{\rm e}$.

The parameters used for the calculations in this paper are summarized in Table~\ref{tab:parameters}.
% : layer thickness $d_{\rm 2D} = 0.626$~nm  \cite{rasmussen2015computational}, single particle band gap ${\varepsilon}_g = 2.48$~eV \cite{rasmussen2015computational}, effective electron mass $m_{\rm e} = 0.43~m_0$ \cite{kormanyos2015k}, effective hole mass $m_{\rm h} = 0.54~m_0$ \cite{kormanyos2015k}, valence-conduction band momentum matrix element $\gamma = 0.222$~eV~nm \cite{kormanyos2015k} and in-plane dielectric constant $\epsilon_\perp = 12.8$ \cite{kumar2012tunable}.

The phonon-induced dephasing rate $\gamma^\text{x}$ is calculated according to the methods given in Ref.~\cite{selig2016excitonic}, without self-consistent inclusion of radiative broadening, because this is contained in the interaction with the quantized electromagnetic field.

In the numerical representation of the equations of motion, the dynamical variables such as $\ev*{a^\dagger}$ and $\ev*{P^\dagger}$ are expressed in surface-density units $\ev*{a^\dagger}/\sqrt{S}$ and $\ev*{P^\dagger}/\sqrt{S}$. This ensures that all coupling coefficients and scattering matrices in the equations of motion are independent of the quantization surface area $S$. Only the input-field driving term contains explicit reference to $S$, when converting the driving power $\mathcal{P}_{\rm in}$ to surface-density units $\mathcal{P}_{\rm in}\rightarrow\mathcal{P}_{\rm in}/S$. For the numerical calculations, $S$ is taken as the laser spot area (see Table~\ref{tab:parameters}).

\rev{In the numerical representation of matrix elements and dynamical variables that depend on wavevectors, we exploit the rotational symmetry and express these quantities in their angle-averaged form as described in detail in Ref.~\onlinecite{takayama2002T}. The complete set of exciton wavefunctions is calculated as the solutions to the angle-averaged Wannier equation (see Ref.~\onlinecite{denning2022efficient} for details), which is solved in momentum space using a polar coordinate system with 4000 points in the radial direction and 4000 points in the angular direction. As discussed in Sec.~\ref{sec:time-evolution}, only the lowest-energy exciton state is kept, and the higher-lying wave functions are discarded. To construct the  matrix elements for the biexcitonic eigenvalue equation (the left-hand-side of of Eq.~\eqref{eq:bix-eigenvalue}) the 2D wavevector-summations in Eq.~\eqref{eq:two-momentum-coulomb-element} must be evaluated for two 2D wave vectors, which presents a significant numerical challenge. Therefore integration in polar coordinates with Gaussian quadratures is used and the matrix elements are calculated for an adjusted, non-equidistant grid of 2D wave vectors, such that the point density is higher for smaller $k$-values and lower for larger $k$-values which are required for convergence. This leads to a matrix with a resolution of 100 $k$-points. Finally, a bicubic spline interpolation is applied to the resulting matrix to form a regular $k$-grid with $100\times100$ $k$-space points and the eigenvalues and eigenvectors are calculated.}

\rev{In the evaluation of some of the matrix elements in Eq.~\eqref{eq:exciton-eom-coefficients}, GPU-based parallelization is used to accelerate the momentum summations using the \texttt{pytorch} library in Python. This makes the evaluation several orders of magnitude faster than a corresponding single-CPU evaluation.}

\rev{The equations of motion Eq.~\eqref{eq:eom} are solved using the DOP853 adaptive 8th order Runge-Kutta algorithm as implemented in the \texttt{scipy.integrate} library in Python. The matrix inversion required to solve the frequency-space Heisenberg-Langevin equation Eq.~\eqref{eq:psi-H-L-1} is carried out using the Python library \texttt{numpy.linalg}. Specifically, the inversion of Eq.~\eqref{eq:inverse-GF} is carried out simultaneously over the dimensions $\zeta,\;m$ and the four vector dimensions of the fluctuation vector in Eq.~\eqref{eq:psi-definition}.}

\begin{table}[b!]
\begin{tabular}{l|c}
Quantity & Value \\
\hline
Exciton-photon coupling $\Omega_0$ & 20 meV \\
Cavity effective refr. index (hBN) $\bar{n}$ & 2.1 \\
Cavity outcoupling rate $\gamma^{\rm p}$ & 9 meV \\
Exciton dephasing rate $\gamma^{\rm x}$ (30 K) & 0.8 meV \\
Pump spot area & 9 $\mathrm{\:\mu m^2}$ \\
Pump power & 5 mW \\
Effective electron mass $m_{\rm e}$ & 0.43 $m_0$~\cite{kormanyos2015k} \\
Effective hole mass $m_{\rm h}$ & 0.54 $m_0$~\cite{kormanyos2015k} \\
In-plane dielectric constant $\epsilon_\perp$ & 12.8~\cite{kumar2012tunable} \\
Single-particle bandgap $E^{\rm c}_0-E^{\rm v}_0$ & 2.48 eV~\cite{rasmussen2015computational} \\
Monolayer thickness $d_{\rm 2D}$ &  0.626~nm~\cite{rasmussen2015computational} \\
Interlayer air gap $h_{\rm int}$ & 0.3 nm \\
Dielectric constant of encapsulating material (hBN) $\varepsilon_{\rm e}$ & 4.5 \\
Val.-cond. band momentum matrix element $\gamma$ & 0.222 eV nm~\cite{kormanyos2015k}
\\
\hline
\end{tabular}
\caption{Parameters used for numerical calculations.}
\label{tab:parameters}
\end{table}

\end{document}